\documentstyle[seceq,epsf,preprint]{ptptex}
\makeatletter
\def\bfm#1{\mbox{\boldmath $#1$}}

\def\lesssim{\mathrel{\mathpalette\vereq<}}
\def\vereq#1#2{\lower3pt\vbox{\baselineskip1.5pt \lineskip1.5pt
\ialign{$\m@th#1\hfill##\hfil$\crcr#2\crcr\sim\crcr}}}
\def\gtrsim{\mathrel{\mathpalette\vereq>}}
\def\alt{\lesssim}
\def\agt{\gtrsim}
\preprintnumber[2.6cm]{\large YITP-03-33}
\makeatother
\title{Color Superconductivity in Schwinger-Dyson Approach\\
				-- {\it Strange Quark Mass and Color-Flavor Unlocking Line} --}
\author{Hiroaki {\sc Abuki}$^{(a,b)}$\footnote{\tt abuki@yukawa.kyoto-u.ac.jp}}
\inst{$(a)$~Yukawa Institute for Theoretical Physics, Kyoto 606-8502, JAPAN\\
		 $(b)$~Department of Physics, University of Tokyo, Tokyo 113-0033, JAPAN}
\abst{%
Phase structure and phase transitions in dense QCD are studied using 
		the Schwinger-Dyson (SD) method in the improved ladder approximation.
We construct the Cornwall-Jackiw-Tomboulis (CJT) effective potentials at finite temperature 
		for two types of pairing ansatz, namely the Color-Flavor locking (CFL) state and the two 
		flavor superconducting (2SC) state. 
Strong coupling effects at low densities, such as the off-Fermi surface and
		antiquarks contribution to the pairing gap due to the large effective coupling, 
		make the gap, critical temperature and those ratio deviate from those weak coupling values.
Nevertheless, the ratio of the physical quantity for the CFL and that for the 2SC 
		does not differ so much from 
		the weak coupling value, as long as the pairing interaction is taken to be the same for the both cases.
As a consequence, the CFL state always dominates over the 2SC state and the critical temperatures
		to the quark-gluon plasma (QGP) phase from both states coincides in the chiral limit.
The energy gain in the CFL state relative to the 2SC state gets smaller towards the critical
		line dividing $(\mu, T)$ plane into the QGP and CFL phases, and thus small perturbations can 
		remove the degeneracy of these critical lines.
As one of such perturbations, the effect of the strange quark mass $m_s$ on the quark-pairing is examined.
In particular,  using a simple kinematical criterion,  we discuss the behaviour of the ``{\it color-flavor unlocking line}''
		on which the CFL phase turns into the 2SC phase, against the variation of $m_s$.}
\begin{document}
\setcounter{secnumdepth}{3}
\newcommand{\beq}{\begin{eqnarray}}
\newcommand{\eeq}{\end{eqnarray}}
\newcommand{\e}{\epsilon}
\newcommand{\ds}[1]{#1 \hspace{-5.2pt}/} 
\newcommand{\Ds}[1]{#1 \hspace{-1ex}/} 
\newcommand{\isum}{\int\hspace{-14.0pt}\sum} 
\newcommand{\isums}{\int\hspace{-10.5pt}\sum} 
\newcommand{\bra}[1]{\left\langle\left. #1 \right\vert\right.}
\newcommand{\ket}[1]{\left.\left\vert #1 \right. \right\rangle}
\newcommand{\braket}[1]{\left\langle #1\right\rangle}
\maketitle
\section{Introduction}
It is believed that the QCD ground state changes against the variation of the external conditions, 
		such as the baryon density, the isospin density and the strange quark mass.
In particular, the color superconducting phases at high baryon density have attracted much interest 
	in high density QCD.
The asymptotic freedom of the QCD \cite{asymp_1,asymp_2} leads to the smaller effective coupling 
		at higher densities~:~%
This gives us the conjecture of deconfined quark matter \cite{Collins_Perry} 
		in which quarks and gluons constitute the active degrees of freedom 
		and are weakly perturbed by the small effective coupling.
Such weakly perturbed deconfined matter might be realized in the core of the neutron stars, 
		or in possible quark stars.
However, if the temperature of such deconfined system is low enough, 
	the residual attractive force between the quarks in the color anti-triplet channel, which is supplied
		by either one gluon exchange (OGE) or the instanton induced interaction, leads to the instability of the Fermi surface. 
This instability causes the non-perturbative generation of the dynamical Majorana mass gap near the Fermi surface,
		which is called ``color superconductivity''.
Along this line, Bailin and Love made an extensive analysis on possible color superconductivities in the 
		quark matter. \cite{BL_84}.

In recent years, there has been great developments in our understanding of the pairing phenomenology
at high density. 
The discovery of ``{\it Color-Flavor Locking}'' (CFL) state \cite{ARW_98} at extremely high density is one of the examples. 
CFL state have two types of  gap in the chiral limit, namely, $\Delta_1$ for the color-flavor singlet excitation, 
		and $\Delta_8$ for the octet excitations.
On the other hand, the 2-flavor superconducting (2SC) state have only one gap $\Delta$, and 
		4 quarks out of 9 obtain this gap.
Rich physics contained in the CFL phase, as well as in the 2-flavor color superconductivity (2SC) phase, 
		has been revealed \cite{Review,Rischke-Review}. 

In our present understanding of the high density QCD, 
		pairing patterns other than the CFL would hardly take place at high density limit \cite{Evans,Hong2}.
It is thought that only in the presence of the finite strange quark mass $m_s\gg m_{u,d}$, the 2SC pairing 
		appears in the low density region in the phase diagram.
The color-flavor unlocking phase transition from the CFL state to the 2SC state at zero temperature have 
		been studied using the Nambu-Jona-Lasinio (NJL) type model \cite{unlock_SW,unlock_AR}. 
This transition occurs when the strange quark mass $m_s$ at some fixed chemical potential $\mu$ gets larger 
		than the critical mass $m_s^c(\mu)$, or equivalently when the chemical potential at some fixed $m_s$ 
		gets smaller than at some critical chemical potential $\mu_c(m_s)$.
Analyses in Refs \cite{unlock_SW,unlock_AR} indicate that a simple kinematical unlocking criterion 
		$m_s^{c2}/4\mu_c\sim |\Delta_8({\mu_c})|_{m_s=0}$ is quite a good guide for the unlocking mass and 
		chemical potential.
This criterion is based on the observation that the unlocking transition would occur at the point where the 
		Fermi momentum for strange quark differs from those for light flavors by the smallest gap $|\Delta_8|$ 
		for the unperturbed ($m_s=0$) CFL state. 
In Ref.~\cite{buballa}, the unlocking (locking) transition at finite temperature is analysed in the NJL model. 

In the chiral limit, weak coupling analyses of the gap equations reveal the CFL dominance 
		over the 2SC at zero temperature and the coincidence of critical temperatures to the QGP 
		phase from the 2SC state and from the CFL state.
The former is attributed to the subtle competition%
		\footnote{The condensation energy in the CFL phase is about twice of that in the 2SC phase.
							If the ratios of the 2SC gap and the CFL gaps, $\Delta/\Delta_{1,8}$ get
						   larger than those weak coupling values by a factor $\sqrt{2}\cong 1.4$ due to
							strong coupling effects, then these two phases may get nearly 
							degenerate with each other in the energy density.}
		between the number of degrees 
		of freedom participating in pairing correlation, and the magnitude of the gaps \cite{Schafer1999}. 
The latter is due to two facts~:~One is the rapid vanishing of the pairing in the color sextet channel 
		towards the critical temperature $T_c$. The other is the disappearance of the nonlinearity of the gap 
		equation in the diquark condensate near $T_c$, and which is guaranteed if the transition is 
		of second order.
However, in the low, and physically interesting density regime, various strong coupling effects
		such as the participation of off-Fermi surface degrees of freedom in the formation of the gap 
		due to the large effective coupling constant \cite{itakura}.
Also the pairing in the color symmetric channel may get relevant towards low densities,
		although it is suppressed by one coupling constant relative to the anti-trriplet 
		one in the weak coupling regime \cite{Schafer1999}.
These strong coupling effects might cause a significant modification of these pictures 
		of the CFL dominance at zero temperature and of the phase transitions. 
In addition, critical temperatures to the quark-gluon plasma (QGP) phase from the CFL and from the 2SC start 
		to deviate towards low densities, although these quantities are shown to be the same in the weak coupling limit \cite{Brown}.
If this is the case, the 2SC phase may show up at high temperature region.%
		\footnote{In ${\rm He}^{3}$ liquid system, the Anderson-Morel (ABM) state 
		\cite{ABMstate} shows up in the high temperature and high pressure region due to strong coupling effects,
		even though the Balian-Wertharmer (BW) state \cite{BWstate} has larger condensation energy in the weak coupling analysis \cite{ABMstate}.}

For the reasons mentioned above, even in the chiral limit, it would be interesting to study the pairing 
		phenomena in dense QCD using some model having strong coupling effects at low density side.
For this purpose, we examine in this paper, the Schwinger-Dyson (SD) approach in the improved
		ladder approximation to the color superconducting phenomena. \cite{itakura,Iwasaki,Matsuzaki,Takagi}
We also use the associated Cornwall-Jackiw-Tomboulis (CJT) potential to determine which phase of 
		(QGP,~2SC,~CFL) is preferred for given density and temperature $(\mu,T)$ \cite{Rischke-Review}.
This approach, namely the SD equation in the Landau gauge with the improved running coupling constant can 
make the asymptotic behaviour of the mass function at high energy consistent with that 
		from the operator product expansion and renormalization group argument in QCD \cite{Higashijima}.
At the same time, the meson properties on the chiral-broken QCD vacuum, can be reasonably understood 
		by the analysis of pseudo scaler Bethe-Salpeter amplitudes \cite{Higashijima,Aoki}.
Also, this approach leads to the correct dependence of the gap on the coupling constant in the weak coupling 
		perturbative regime.
Thus, we believe the improved SD approach is reliable for a wide region of the quark chemical potential.
Various strong coupling effects, say, the large coupling constant, pairing correlation far away from the Fermi surface, 
		and that for anti-quark channel, all become relevant for realistic densities in this improved SD approach \cite{itakura,Matsuzaki}.
Also, we investigate the effect of the finite strange quark mass on the pairing phenomena, using a simple 
		criterion for the color-flavor unlocking phase transition.

This paper is organized as the following.
Sec.~\ref{sec:2} is devoted to the introduction of our model and the technical details to derive the gap equations, 
	the CJT effective potentials and other physical quantities characterizing the CFL and 2SC states.  
In Sec.~\ref{sec:3}, we present our numerical results.
We discuss the superconducting states at finite $T$ and study the phase transitions to normal quark matter in 
	details at $\mu=1000$~MeV. 
Also, how the 2SC and CFL states at $T=0$ and transition temperatures $T_c$ are modified towards the lower density 
	region by strong coupling effects. 
At the last of the section, we discuss the phase diagram from our model and its dependence on the strange quark mass $m_s$. 
In Sec.~\ref{sec:4}, the summary of this work and perspective for the future work will be given.

\section{Gap equations and CJT potentials for 2SC and CFL states}\label{sec:2}
In this section, we present a general framework
	to derive the gap equation for the given order parameter, 
	focusing on the 2SC and CFL phases. 
Also, we give various physical quantities characterizing the ground state.

\subsection{Gap equations for the 2SC and CFL states}
{\it Nambu-Gor'kov propagators~:~~}Here, we introduce two component quark fields $\Psi=(q,q_c)^t$ in order 
	to express the superconducting gap as a part of the self-energy. In this Nambu-Gor'kov base, 
quark propagators are introduced as
\beq
		i{\bfm{S}}(p_0,\bfm{p})&=&%
				i\left(\matrix{S_{11}(p_0,\bfm{p}) & S_{12}(p_0,\bfm{p})\cr
                        S_{21}(p_0,\bfm{p}) & S_{22}(p_0,\bfm{p})\cr}%
         \right)\nonumber\\
					&=&\int d^4z\,e^{ip_\mu z^\mu}\left(\matrix{
          {\langle Tq(z)\bar{q}(0)\rangle} & 
					{\langle Tq(z)\bar{q}_c(0)\rangle}\cr
        	{\langle Tq_c(z)\bar{q}(0)\rangle} & 
					{\langle T{q}_c(z)\bar{q}_c(0)\rangle}\cr}%
      		\right).\nonumber
\eeq
Here, the subscript $c$ represents the charge conjugation 
	$q^c=C\bar{q}^t$, and $\langle T{\mathcal O}\rangle$ means the ground state expectation value 
of the time ordered product of the composite operator ${\mathcal O}$.
We introduce the full self-energy matrix for the quark field in this base as
\beq
	\bfm{\Sigma}(p_0,\bfm{p})=\left(
                  \matrix{\Sigma_{11}(p_0,\bfm{p}) & \Sigma_{12}(p_0,\bfm{p}) \cr
                          \Sigma_{21}(p_0,\bfm{p}) & \Sigma_{22}(p_0,\bfm{p}) \cr}
                  \right).\nonumber
\eeq
This self-energy gets dynamically generated by the interaction among quarks
and defines the asymptotic quark fields. If the interaction is invariant
under charge conjugation, the relation 
$\Sigma_{22}(p_0,\bfm{p})=-C\Sigma_{11}(-p_0,-\bfm{p})^tC$ 
should hold. If the interaction is hermitian so that the time evolution of the 
system is unitary, the other relation 
$\Sigma_{21}(p_0,\bfm{p})=\gamma_0\Sigma_{12}(p_0,\bfm{p})^\dagger\gamma_0$ 
also holds. As a result, four components in ${\bfm{S}}(p)$ are not independent. 
\beq
      &&iS_{22}(t,\bfm{z})=-C(iS_{11}(-t,-\bfm{z}))^tC ,\quad %
				S_{22}(p_0,\bfm{p})=-CS_{11}(-p_0,-\bfm{p})^tC,\\
      &&iS_{21}(t,\bfm{z})=-\gamma_0(iS_{12}(-t,-\bfm{z}))^\dagger\gamma_0,%
			\quad S_{21}(p_0,\bfm{p})=\gamma_0S_{12}(p_0,\bfm{p})^\dagger\gamma_0.
\eeq
Our main purpose is to investigate the density region where the chiral symmetry is already recovered. 
Therefore, we assume that the diagonal elements of the self-energy, namely,
		Dirac mass function, the deviation of the quark wave function renormalization 
		from $1$, and the radiative correction to the chemical potential are all zero.
We postulate the following forms for the off-diagonal elements 
\cite{BL_84,Pisarski-Rischke2,Pisarski-Rischke,Schaefer,Hong},
\beq
	    \Sigma_{12}(p_0,\bfm{p})&=&\gamma_5(\Lambda_+(\hat{p})\hat{\Delta}_-(p_0,\bfm{p})%
			+\hat{\Lambda}_-(\hat{p})\hat{\Delta}_+(p_0,\bfm{p})),\label{eq:02}\\
    \Sigma_{21}(p_0,\bfm{p})&=&-\gamma_5(\Lambda_-(\hat{p})\hat{\Delta}_-^\dagger(p_0,\bfm{p})%
			+\Lambda_+(\hat{p})\hat{\Delta}_+^\dagger(p_0,\bfm{p})).\label{eq:03}
\eeq
$\Lambda_{\pm}(\hat{p})$ in Eqs.~(\ref{eq:02}) and (\ref{eq:03}) are the 
projection matrices for positive and negative energy states in the massless limit~:~%
		$\Lambda_{\pm}(\hat{p})={(1\pm\bfm{\alpha}\cdot\hat{p})}/{2}$, where 
		$\bfm{\alpha}\equiv\gamma_0\bfm{\gamma}$ and $\hat{p}\equiv\bfm{p}/|\bfm{p}|$.
$\hat{\Delta}_{\pm}(p_0,\bfm{p})$ are the gap functions and are still matrices
		in the color and flavor spaces.
Eigenvalues of $\hat{\Delta}_\pm$ turn out to be gap functions in quasi-quarks ($+$) and 
		quasi-antiquark ($-$) excitations. 
$\gamma_5$ in Eqs.~(\ref{eq:02}) and (\ref{eq:03}) guarantees that the pairing
	occurs in positive parity channel. 
Although the OGE does not distinguish the positive and negative parity channels, 
	the instanton induced interaction which may be relevant at low densities, 
	prefers the pairing in the positive parity channel \cite{RSSV_98,ARW_98b}.
The Dirac structures of Eqs.~(\ref{eq:02}) and (\ref{eq:03}) are of 
	the most general pairing form for the $J=0^+$ with aligned chirality \cite{Pisarski-Rischke2}.%
	\footnote{There is no Cooper instability in the mixed chirality $q_L(\bfm{q})q_R(-\bfm{q})$ channel \cite{SW_3}, 
	thus the condensate for this channel gets suppressed by the factor $m_q/\mu$.}
For the moment, we proceed without specific ansatz for the color-flavor structure. 
The inverse of the quark propagator can be expressed as
\beq
   \bfm{S}^{-1}(p_0,\bfm{p})=\left(
                   \matrix{\ds{p}+\ds{\mu}   & -\Sigma_{12}(p_0,\bfm{p})\cr
                   -\Sigma_{21}(p_0,\bfm{p}) & \ds{p}-\ds{\mu}\cr}\right).
\eeq
Here $\ds{\mu}=\mu\gamma_0$ and $\mu$ is the chemical potential for net quark.
By taking the inverse of this expression, we get
\beq
   S_{11}(p_0,\bfm{p})&=&+\frac{\ds{p}_+}{2p}\frac{(p_0+E_+)}{p_0^2-E_+^2-\hat{\Delta}_+^2}
               -\frac{\ds{p}_-}{2p}\frac{(p_0-E_-)}{p_0^2-E_-^2-\hat{\Delta}_-^2},\\
   S_{12}(p_0,\bfm{p})&=&-\gamma_5\left(
                \Lambda_+(\hat{p})%
                \frac{\hat{\Delta}_+}{p_0^2-E_+^2-\hat{\Delta}_+^2}
               +\Lambda_-(\hat{p})%
                \frac{\hat{\Delta}_-}{p_0^2-E_-^2-\hat{\Delta}_-^2}\right),
\eeq
where we have defined $E_\pm=p\mp\mu$ and $p=|\bfm{p}|$, 
the bare dispersions for quark and anti-qaurk in the system with the chemical 
potential $\mu$. $p_{\pm}$ are defined as the forward and off-forward light cone vectors,
	$p_{\pm}=(1,\pm\hat{p})$, so that $\ds{p}_{\pm}=\gamma_0\mp\bfm{\gamma}\cdot\hat{p}$.

{\it Gap equation as a self-consistentcy condition for the self-energy~:~~}The 
gap equations, the self-consistency conditions for $\hat{\Delta}_\pm(p_0,p)$ 
up to one loop order, can be expressed as the off-diagonal sector of the following
	Schwinger-Dyson equation in the Nambu-Gor'kov base.
\beq
	  \bfm{\Sigma}(p_0,\bfm{p})%
		&=&ig^2\int\!\!\!\frac{d^4q}{(2\pi)^4} {\bfm{\Gamma}}_a^\mu\,%
		\bfm{S}(q_0,\bfm{q})\bfm{\Gamma}_b^\nu(q,p)\,D^{ab}_{\mu\nu}(q_0-p_0,\bfm{q}-\bfm{p}).
\eeq
For simplicity, we use the bare vertex instead of the full vertex throughout this paper.
\beq
		\bfm{\Gamma}_{a}^\mu\equiv\left(\matrix{\Gamma_a^\mu & 0\cr
 											              0 & \bar{\Gamma}_a^\mu}\right)%
              =\left(\matrix{t_a\gamma_\mu & 0\cr
 											              0 & -t_a^t\gamma_\mu}\right).
\eeq
$t_a$ is defined as $\lambda_a/2$, where $\{\lambda_a\}$ 
	are Gell-Mann matrices. 
The vertex $\bar{\Gamma}_{A}=-t_A^t\gamma_\mu$ originates from the 
	relation $\bar{q}\Gamma_Aq=-\bar{q}_cC\Gamma_A^tC{q}_c$.
Extra minus sign purely comes from the Fermi statistics. 
$D^{ab}_{\mu\nu}(q_0-p_0,\bfm{q}-\bfm{p})\equiv\delta^{ab}{\mathcal D}_{\mu\nu}%
	(q_0-p_0,\bfm{q}-\bfm{p})$ is the gluon propagator in the medium, and 
	the detailed structure of the function ${\mathcal D}_{\mu\nu}(q_0-p_0,\bfm{q}-\bfm{p})$ 
	which we use in this paper will be specified latter.

Through the positive and negative energy projection procedures, we obtain
the following gap equations,
\beq
		\hat{\Delta}_\pm(p_0,\bfm{p})&=&ig^2\!\!\int\!\!\!\frac{d^4q}{(2\pi)^4}%
				\!\left(
 				c_{\mu\nu}(p_\mp,q_+)\frac{t_A^t\hat{\Delta}_+(q_0,\bfm{q})t_A}{q_0^2-E_+^2-\hat{\Delta}_+^2}\right.\nonumber\\
			  & &\qquad\qquad\left.%
				+c_{\mu\nu}(p_\mp,q_-)\frac{t_A^t\hat{\Delta}_-(q_0,\bfm{q})t_A}{q_0^2-E_-^2-\hat{\Delta}_-^2}%
				\right)\!{\cal D}_{\mu\nu}(q_0-p_0,\bfm{q}-\bfm{p}),
\eeq
where we have defined $c_{\mu\nu}(p,q)$ by 
	$c_{\mu\nu}(p_\pm,q_\pm)\equiv-(1/2){\rm tr}\left(\Lambda_\pm(\hat{p})\gamma_\mu\Lambda_\pm(\hat{q})\gamma_\nu\right)$

{\it Color-flavor structure of the gap matrix in the 2-flavor case~:~~}In 
the 2 flavor case, the gap matrix in the color-flavor space is specified as \cite{BL_84}:
\beq
	\hat{\Delta}^{\rm 2SC}_\pm(p_0,\bfm{p})=(\tau_2\times\lambda_2)^{ab}_{ij}\Delta_\pm(p_0,\bfm{p}).
\eeq
Because $\tau_2\lambda_2$ is rank 4 matrix in the color-flavor space, 5 quarks $(gu,gd,gs,rs,bs)$ 
corresponding to five eigenvectors with zero eigenvalue of the gap matrix remain gapless.
These quarks contain the quantum number {\it strangeness} or {\it green color}, 
	which does not take part in pairing.
Other 4 modes belonging to the unbroken $SU(2)_c\times SU(2)_f$ doublets 
$[{\bf 2}_{(r,b)},{\bf 2}_{(u,d)}]$ obtain the finite gaps $\Delta_\pm$.
The spectral density matrix $\bfm{\rho}$ for the 2SC state 
	can be extracted from the retarded propagator $S^R_{\rm 2SC}$. 
Its off-diagonal element $\rho_{12}$ can be written as
\beq
	   {\rho}_{12}^{\rm 2SC}(p_0,\bfm{p})&=&-(\tau_2\lambda_2)^{ab}_{ij}%
						\big[\gamma_5\Lambda_+(\hat{p})\Delta_+2\pi\epsilon(p_0)\delta(p_0^2-E_+^2-\Delta_+^2)\nonumber\\%
							& &\qquad\qquad-\gamma_5\Lambda_-(\hat{p})\Delta_-2\pi\epsilon(p_0)\delta(p_0^2-E_-^2-\Delta_-^2)\big].
\eeq
The conditions $p_0^2-E_\pm^2-\Delta_\pm(p_0,\bfm{p})^2=0$ define the quasi-quark 
and quasi-antiquark dispersions. 

{\it Color-flavor structure of the gap matrix in the 3-flavor case~:~~}The 
color-flavor locked state is defined as \cite{ARW_98},
\beq
	\hat{\Delta}^{\rm CFL}_\pm(p_0,p)&=&(P_1)^{ab}_{ij}\Delta^1_{\pm}(p_0,p)%
												+(\bar{P}_8)^{ab}_{ij}\Delta^8_{\pm}(p_0,p),\label{eq:ODCFL}\\
										 &=&\frac{1}{N_c}\delta^a_i\delta^b_j\Delta^1_{\pm}%
											 +\left(\delta^a_j\delta^b_i-\frac{1}{N_c}\delta^a_i\delta^b_j\right)\Delta^8_{\pm}.
\eeq
The matrix $\hat{\Delta}$ is full-ranked in the color-flavor space, which results in generating 
		gaps in all quasi-quark excitations.%
\footnote{The CFL states is similar to the Balian-Wertharmer (BW) state \cite{BWstate} 
		in the following sense.
	The BW state is characterized by the unit gap matrix in the angular momentum and 
		spin spaces, and has a gap on the whole Fermi surface.
	On the other hand, the CFL state is also a diagonal unit matrix 
		in the color-flavor $\bar{\bf 3}_c\times\bar{\bf 3}_f$ space, 
		and this leads non-zero gap for all nine quarks in color and flavor space.
	The 2SC state is much like to Anderson-Morel (ABM) state \cite{ABMstate}, 
		or to polar state, which is not diagonal and isotropic state, 
		and has gapless point or line on the Fermi surface.}
In the chiral limit, the CFL state has two-types of gap, corresponding to
		two eigenvalues in this matrix, one being $\Delta_1$ and the other being $\Delta_8$. 
$P_{1(8)}$ is the projection to the asymptotic quark which transforms as singlet (octet) under the 
		simultaneous inverse rotation $SU(3)_{C+V}\in SU(3)_c\times SU(3)_f$ in the color-flavor mixed space.%
\footnote{$\bar{P}_8$ is a square root of the octet projection matrix 
		$(P_8)^{ab}_{ij}=\delta^{ab}\delta_{ij}-\delta^{a}_{i}\delta^{b}_{j}/N_c$.} 
The CFL order parameter (\ref{eq:ODCFL}) can be divided into states in the color multiplets 
	$(\bar{\bf3}_c,{\bf6}_c)$ as
\beq
		&\hat{\Delta}^{\rm CFL}=\epsilon^{Iab}\epsilon_{Iij}\Delta_A+(\delta^a_i\delta^b_j+\delta^a_j\delta^b_i)\Delta_S&\\
	  &\Delta_A=\frac{1}{6}\Delta^1-\frac{2}{3}\Delta^8,\quad%
			\Delta_S=\frac{1}{6}\Delta^1+\frac{1}{3}\Delta^8,&\label{eq:dec2}
\eeq
where $N_c=3$ is taken. 
$\Delta_A$ represents self-energy in the color anti-symmetric and flavor antisymmetric channel, 
	where the attraction works.
However, it turns out that the nonzero $\Delta_S$ gets induced in a non-perturbative 
	way as a result of the self-consistent treatment of the gap equation \cite{ARW_98}. 
One can easily extract the spectral density for off-diagonal propagator,
\beq
	{\rho}_{12}^{\rm CFL}(p_0,\bfm{p})&=&-(P_1)^{ab}_{ij}\big[\gamma_5\Lambda_+\Delta_{1+}2\pi\epsilon(p_0)\delta(p_0^2-E_{+}^2-\Delta_{1+}^2)\nonumber\\
							& &\qquad\qquad+\gamma_5\Lambda_-\Delta_{1-}2\pi\epsilon(p_0)\delta(p_0^2-E_{-}^2-\Delta_{1-}^2)\big]\nonumber\\
							& &-(\bar{P}_8)^{ab}_{ij}\big[\gamma_5\Lambda_+\Delta_{8+}2\pi\epsilon(p_0)\delta(p_0^2-E_{+}^2-\Delta_{8+}^2)\nonumber\\
							& &\qquad\qquad+\gamma_5\Lambda_-\Delta_{8-}2\pi\epsilon(p_0)\delta(p_0^2-E_{-}^2-\Delta_{8-}^2)\big].
\eeq
The condition $p_0^2-E^2-\Delta_{1(8)}(p_0,\bfm{p})=0$ defines the dispersions 
for the $SU(3)_{C+V}$ singlet (octet) quasi-quark excitations in the CFL state.

The Feynman propagator for $T\neq 0$ can be obtained from above spectral densities,
\beq
		S_{F12}(p_0,\bfm{p})&=&%
			-\frac{\gamma_5\Lambda_+\hat{\Delta}_+(p_0,\bfm{p})}{p_0^2-(|\bfm{p}|-\mu)^2-\hat{\Delta}_+^2+i\eta}%
			-\frac{\gamma_5\Lambda_-\hat{\Delta}_-(p_0,\bfm{p})}{p_0^2-(|\bfm{p}|+\mu)^2-\hat{\Delta}_-^2+i\eta}\nonumber\\
			& &+in_F(|p_0|)\epsilon(p_0){\rho}_{12}(p_0,\bfm{p};\mu).
\eeq
$\hat{\Delta}$ and $\rho_{12}$ is color-flavor matrix each for the 2SC and the CFL.
$n_F$ is the thermal Fermi distribution function defined by
\beq
	n_F(|p_0|)=\frac{1}{1+e^{|p_0|/T}}.
\eeq
This produces the Pauli-blocking term in the final gap equation
by thermal on-shell quarks at finite $T$, and suppresses the phase space
for pairing correlation and brings about the phase transition to normal phase at some critical temperature.

{\it Gap equations for the 2SC and CFL states~:~~}Now 
we simplify the gap equation through the vertex decomposition.
In the case of the 2SC, we use the following equation,
\beq
		\sum_{A=1}^{8}t_A^t\lambda_2t_A=-\frac{N_c+1}{2N_c}\lambda_2.
\eeq 
In the case of the CFL, we use the following color-flavor algebra,
\beq
		\sum_{A=1}^{8}t_A^t(P_1)t_A=\frac{1}{2N_c}(\bar{P}_8),\quad%
		\sum_{A=1}^{8}t_A^t(\bar{P}_8)t_A=\frac{N_c^2-1}{2N_c}(P_1)-\frac{1}{N_c}(\bar{P}_8).
\eeq
These identities can be derived by the Fierz identity~:~%
$$\sum_{A=1}^{8}(\lambda_A)_{ab}(\lambda_A)_{cd}=-(2/N_c)\delta_{ab}\delta_{cd}+2\delta_{ad}\delta_{bc}.$$
Using these identities, we obtain
\beq
	\frac{2N_c}{N_c+1}\Delta_+=K^+\left[\Delta_+,\Delta_-\right]&,&\,\,
   \frac{2N_c}{N_c+1}\Delta_-=K^-\left[\Delta_+,\Delta_-\right],\label{eq:001}
\eeq
for the 2SC, and
\beq
	   \!\!\!\!\!\!\!\!\!\!\!\frac{2N_c}{N_c^2-1}\Delta_{1+}=-K^+\left[\Delta_{8+},\Delta_{8-}\right]&,&\,\,
		2N_c\Delta_{8+}+\frac{4N_c}{N_c^2-1}\Delta_{1+}=-K^+\left[\Delta_{1+},\Delta_{1-}\right],\label{eq:003}\\
		\!\!\!\!\!\!\!\!\!\!\!\frac{2N_c}{N_c^2-1}\Delta_{1-}=-K^-\left[\Delta_{8+},\Delta_{8-}\right]&,&\,\,
	2N_c\Delta_{8-}+\frac{4N_c}{N_c^2-1}\Delta_{1-}=-K^-\left[\Delta_{1+},\Delta_{1-}\right].\label{eq:002}
\eeq
for the CFL. 
We define the kernel integral for OGE as:
\beq
		\!\!\!\!\!\!K^{\alpha}\left[\Delta_+,\Delta_-\right]&=&%
			-ig^2\int\frac{d^4q}{(2\pi)^4}\Big\{%
				S_{F}^+(q;\Delta_+)c_{\mu\nu}(p_{\bar{\alpha}},q_+)\nonumber\\
				& &\qquad\qquad\qquad+S_{F}^-(q;\Delta_-)c_{\mu\nu}(p_{\bar{\alpha}},q_-)%
				\Big\}{\cal D}_{\mu\nu}(q_0-p_0,\bfm{q}-\bfm{p}),
\eeq 
with definition
\beq
		S^\pm_F(q;\Delta_\pm)&=&\frac{\Delta_\pm}{q_0^2-(|\bfm{q}|\mp\mu)^2-\Delta_\pm^2+i\eta}\nonumber\\%
						& &\quad+2\pi in_F(|q_0|)\delta\left[q_0^2-(|\bfm{q}|\mp\mu)^2-\Delta_\pm^2\right].
\eeq
Index $\alpha$ takes $+$ and $-$, and we define $\bar{\alpha}\equiv-\alpha$.
These gap equations are the generalization of those obtained in 
Ref.~\cite{ARW_98} to the non-local interaction in arbitrary $N_c$ with 
the antiquark gap.

{\it Gluon propagator in the quark Fermi liquid~:~~}We 
use the quasi-static approximation of the hard dense loop
propagator in the Landau gauge, as it is taken in Refs.~\cite{itakura,Pisarski-Rischke,Iida}.
\beq
	{\cal D}_{\mu\nu}(p_0,\bfm{p})=-\Delta_T(p_0,\bfm{p})P^T_{\mu\nu}%
														-\Delta_L(p_0,\bfm{p})P^L_{\mu\nu},
\eeq
with the transverse and longitudinal projection matrices,
\beq
	P^T_{ij}=\delta_{ij}-\hat{p}_i\hat{p}_j,\quad
	P^T_{0i}=P^T_{i0}=P^T_{00}=0,\quad
	P^L_{\mu\nu}=-g_{\mu\nu}+\frac{p_\mu p_\nu}{p_0^2-p^2}-P^T_{\mu\nu},
\eeq
and the corresponding amplitudes
\beq
		\Delta_T(p_0,\bfm{p})=\frac{1}{p^2+i\theta(p-p_0)(\pi m_D^2/4)|p_0|/p},\quad%
		\Delta_L(p_0,\bfm{p})=\frac{1}{p^2+m_D^2}.
\eeq
$m_D^2=N_fg^2\mu^2/2\pi^2$ is the Debye screening mass in the quark-gluon plasma.
In magnetic sector, we have to take into account the subleading correction from 
the dynamical sector up to linear in $p_0/p$ in order to avoid the collinear singularity.%
\footnote{The quasi-static approximation can be justified if the ratio of the obtained gap 
	to the Fermi surface $(\Delta/\mu)$ is small enough to neglect $(\Delta/\mu)^2$.
   The quasi-static approximation is going well in our model, because it will turn out
	$\Delta/\mu\sim 1/10$ even at the lowest density.}
We have neglected the effect of the temperature on the polarization because our main
interest lies in the region $\Delta(\bfm{p})|_{T=0}\ll\mu$. 

{\it Gap equation for quasi-particle on the mass shell~:~~}Now 
we simplify the gap equation by integrating out the frequency
as in Ref.~\cite{itakura} 
because our main interest lies in the gap of quasi-quarks on the mass shell.
In the kernel $K^\alpha$ of the gap 
equations, we have three contributions: quark pole, plasmon pole and 
blanch cut in the gluon continuum.
It has been shown in Ref.~\cite{Pisarski-Rischke} that only the quark 
pole has a significant effect on gap equations at least in the weak coupling 
limit $\Delta(\bfm{p})/\mu\to 0$ at $T=0$. Here, we assume this property 
holds even at finite temperatures and/or in the strong coupling regime.
The resulting gap equations for the 2SC and CFL states under quark pole 
	dominance take the same forms as Eqs.~(\ref{eq:001})$\sim$(\ref{eq:002}) 
	under replacement of the kernels $K^{\alpha}$ by :
\beq
	\bfm{K}^{\pm}\left[\bfm{p};\Delta_\pm\right]%
		&\equiv&\int\frac{d\bfm{q}}{(2\pi)^3}{\cal D}_{\pm+}(\bfm{p},\bfm{q})%
				\frac{\Delta_+(\bfm{q})}{2\epsilon_{+}(\bfm{q})}%
				\Big[1-2n_F\big[\epsilon_{+}(\bfm{q})\big]\Big]\nonumber\\
		&			  &+\int\frac{d\bfm{q}}{(2\pi)^3}{\cal D}_{\pm-}(\bfm{p},\bfm{q})%
				\frac{\Delta_-(\bfm{q})}{2\epsilon_{-}(\bfm{q})}%
				\Big[1-2n_F\big[\epsilon_{-}(\bfm{q})\big]\Big],\label{zzz}
\eeq
where $\e_\pm(\bfm{p})=\sqrt{(p\mp\mu)^2+\Delta_\pm(\bfm{p})^2}$, and $\Delta_\pm(\bfm{p})$ are
	gaps on the quasi-quark and quasi-antiquark on the mass shell. 
The effective interaction ${\cal D}_{\alpha\beta}(\bfm{q},\bfm{p})$ 
is constructed by the following electric and magnetic interactions,
\beq
	{\cal D}_{\alpha\beta}(\bfm{p},\bfm{q})%
	={\cal D}^{\rm E}_{\alpha\beta}(\bfm{p},\bfm{q})%
		+2{\cal D}^{\rm M}_{\alpha\beta}(\bfm{p},\bfm{q}),\label{www}
\eeq
with 

\beq
 {\cal D}_{+\pm}^{\rm E}(\bfm{p},\bfm{q})&=&{\cal D}_{-\mp}^{\rm E}(\bfm{p},\bfm{q})%
		=\frac{g^2(1\pm\hat{p}\cdot\hat{q})/2}{(\bfm{p}-\bfm{q})^2+m_D^2},\nonumber\\
 {\cal D}_{\alpha\pm}^{\rm M}(\bfm{p},\bfm{q})&=&\frac{1}{4}{\rm Re}%
		 \left[\frac{g^2((\bfm{p}-\bfm{q})^2\pm(p-\hat{p}\cdot\bfm{q})(q-\hat{q}\cdot\bfm{p}))}%
				{(\bfm{p}-\bfm{q})^4+i(\pi m_D^2/4)|\epsilon_\alpha(\bfm{p})-\epsilon_\pm(\bfm{q})||\bfm{p}-\bfm{q}|}\right.\nonumber\\
				& &\qquad\left.+\frac{g^2((\bfm{p}-\bfm{q})^2\pm(p-\hat{p}\cdot\bfm{q})(q-\hat{q}\cdot\bfm{p}))}%
				{(\bfm{p}-\bfm{q})^4+i(\pi m_D^2/4)|\epsilon_\alpha(\bfm{p})+\epsilon_\pm(\bfm{q})||\bfm{p}-\bfm{q}|}\right]+i{\mathcal} O(g^4).\nonumber
\eeq
The Fermi distribution function $n_F$ in Eq.~(\ref{zzz}) represents Pauli blocking 
by thermally excited quasi-quark and quasi-antiquarks.
After integration over angular variables, we obtain the following kernels.
\beq
  &  &\bfm{K}^{\pm}(p;\Delta_\pm)\nonumber\\
  &  &\,\,=+\frac{g^2}{16\pi^2}\int_0^\infty dq\frac{q}{k}%
			\tanh\left(\frac{\epsilon_+}{2T}\right)\frac{\Delta_+(q)}{2\epsilon_+(q)}\biggl\{
			\pm\frac{(p\pm q)^2+m_D^2}{2pq}\ln\biggl(\frac{(p+q)^2+m_D^2}{(p-q)^2+m_D^2}\biggl)\nonumber\\
	& &\quad\,\,+\frac{1}{3}\ln\biggl(\frac{(p+q)^6+M^4(\e_\pm(p)-\e_+(q))^2}{(p-q)^6+M^4(\e_\pm(p)-\e_+(q))^2}%
					\cdot\frac{(p+q)^6+M^4(\e_\pm(p)+\e_+(q))^2}{(p-q)^6+M^4(\e_\pm(p)+\e_+(q))^2}\biggl)\biggl\}\nonumber\\
	& &\quad\,\,+\frac{g^2}{16\pi^2}\int_0^\infty dq\frac{q}{k}%
			\tanh\left(\frac{\epsilon_-}{2T}\right)\frac{\Delta_-(q)}{2\epsilon_-(q)}\biggl\{%
			\mp\frac{(p-q)^2+m_D^2}{2pq}\ln\biggl(\frac{(p+q)^2+m_D^2}{(p-q)^2+m_D^2}\biggl)\nonumber\\
	& &\quad\,\,+\frac{1}{3}\biggl[\ln\biggl(\frac{(p+q)^6+M^4(\e_\pm(p)-\e_-(q))^2}{(p-q)^6+M^4(\e_\pm(p)-\e_-(q))^2}%
			\cdot\frac{(p+q)^6+M^4(\e_\pm(p)+\e_-(q))^2}{(p-q)^6+M^4(\e_\pm(p)+\e_-(q))^2}\biggl)\biggl]\biggl\}\nonumber,
\eeq
where we have defined $M^2\equiv\pi m_D^2/4$.
We have neglected the parts with higher angular momenta in the gap function 
$\Delta(\bfm{q})$.\footnote{%
The gap function $\Delta(\bfm{q})$ can be expanded as
$
	\Delta(\bfm{q})=\Delta(q)+\sum_{l=1}^\infty\Delta^l(q)P_l(\cos\theta)
$.
We assume $\Delta^l\ll\Delta$, because the isotropic gap function
can activate the whole Fermi surface where the infrared Cooper singularity lies.
}
Following Ref.~\cite{itakura}, we replace the coupling $g^2$ with the momentum 
	dependent effective one $\bar{g}^2(p,q)$ by the Higashijima-Miransky prescription \cite{Higashijima}. 
In the improved ladder approximation, $\bar{g}(p,q)^2$ is taken to be
\beq
	\bar{g}^2(p,q)=\frac{16\pi^2}{\beta_0}\frac{1}{\ln((p_{\rm max}^2+p_c^2)/\Lambda^2)},\quad%
	{p_{\rm max}}={\rm max}(p,q),
\eeq
where $\beta_0=(11N_c-2N_f)/3$, $p_c^2$ plays a role of a phenomenological infra-red regulator.
We adopt $\Lambda=400$ MeV and $p_c^2=1.5\Lambda^2$ \cite{Higashijima} for numerical calculations.

\subsection{CJT effective potentials for finite temperature}
The CJT effective potential \cite{CJT} (The Ward-Lattinger effective potential 
in the case of non-relativistic fermion) for fermion propagator is given by,
\beq
	\Gamma\left[\bfm{S},\mu\right]%
		&=&\frac{1}{2}\left\{{\rm TrLog}\left[\bfm{S}\right]%
			-{\rm Tr}\left[\bfm{S}\bfm{S}_0^{-1}\right]+V\left[\bfm{S}\right]\right\}.\label{eq:CJJT}
\eeq
Here, ``TrLog'' should be interpreted in the functional sense. In the momentum space,
$\bfm{S}^{-1}(p_0,\bfm{p})=\bfm{S}_0^{-1}(p_0,\bfm{p})-\bfm{\Sigma}(p_0,\bfm{p})$ with 
$\bfm{S}_0^{-1}(p_0,\bfm{p})={\rm diag.}(\ds{p}+\ds{\mu},\ds{p}-\ds{\mu})$.
The overall factor $1/2$ in the Eq.~(\ref{eq:CJJT}) should be introduced to eliminate 
the artificial degrees of freedom introduced by the Nambu-Gor'kov 2-component fermion. 
$\mu$ represents explicit $\mu$ dependence through the 
bare propagator and the gluon polarization. $V[\bfm{S}]$ is a potential functional
which contains Feynman diagrams of the skeleton self-energy and one quark 
propagator $S$. This is a generating functional of the proper self-energy, 
namely, the functional derivative of the potential is the self-energy, 
$\delta{V[\bfm{S}]}/\delta\bfm{S}=\Sigma[\bfm{S}]$. 
Now we take one loop approximation for $\Sigma[\bfm{S}]$, 
symbolically, written as $\Sigma^{(1)}[\bfm{S}]=g^2\bfm{\Gamma}_0\bfm{S}\bfm{\Gamma} D$.%
\footnote{$\bfm{\Gamma}_0$ and $\bfm{\Gamma}$ means the bare and full vertices
in the Nambu-Gor'kov base.} 
In this case, $\Gamma[\bfm{S}]$ reduces to 
\beq
		\Gamma^{(2)}\left[\bfm{S},\mu\right]%
			&=&\frac{1}{2}\left\{{\rm TrLog}\left[\bfm{S}\right]%
			-{\rm Tr}\left[\bfm{S}\bfm{S}_0^{-1}\right]%
			+\frac{1}{2}{\rm Tr}\left[\bfm{S}\Sigma^{(1)}%
			\left[\bfm{S}\right]\right]\right\}.
\eeq
The saddle point approximation leads to the one loop gap equation : 
		${\delta\Gamma^{(2)}}/{\delta\bfm{S}}%
		=\bfm{S}^{-1}-\bfm{S}_0^{-1}+\bfm{\Sigma}^{(1)}\left[\bfm{S}\right]=0$.
We write the solution of this equation $\bar{\bfm{S}}(g;\mu)$ and 
$\bfm{S}_0^{-1}-\bar{\bfm{S}}^{-1}\equiv\bar{\bfm{\Sigma}}(g;\mu)$.
At the stationary point, we can simplify the effective potential:
\beq
		\Gamma^{(2)}\left[\bar{\bfm{S}}(g;\mu),\mu\right]%
			&=&\frac{1}{2}\left\{{\rm TrLog}\left[\bar{\bfm{S}}\right]%
			   -\frac{1}{2}{\rm Tr}\left[\bar{\bfm{S}}\bar{\bfm{\Sigma}}\right]\right\}.	
\eeq
This is a useful formula in which the coupling and gluon propagator 
do not appear explicitly, but are hidden in $\bar{S}$ and $\bar{\Sigma}$.
Using CJT method described above, the thermodynamic potential (condensation energy) 
density relative to the normal Fermi gas up to 2-loop order is given by
\beq
		\delta\Omega(\mu,T)=\frac{1}{\beta V}\left\{\frac{1}{2}{\rm TrLog}%
							 \left[\bfm{S}\bfm{S}_0^{-1}\right]%
							-\frac{1}{4}{\rm Tr}\big[\bfm{S}\bfm{\Sigma}\big]\right\}.
\eeq
Note that this expression is only valid at the stationary point which can be determined 
by the gap equation for $\bfm{\Sigma}$.
\begin{figure}[tbp]
  \centerline{
  \epsfsize=0.33\textwidth
  \epsfbox{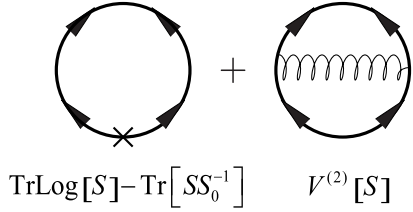}}
 \caption[]%
	{Schematic Feynman graph of 2-loop CJT potential functional.
	$\bfm{S}$-derivative on this potential is removing one solid line,
	which yields one loop gap equation.}
 \label{fig:CJT}
\end{figure}
After an involved but straightforward algebra and Matsubara summation, 
we reach the following expression for the 2SC phase,
\beq
	\frac{\delta\Omega_{\rm 2SC}}{6N_c}%
			&=&\frac{2}{3}\frac{N_c-1}{N_c}\biggl[-T\int\frac{d\bfm{q}}{(2\pi)^3}%
				\left\{\ln\left[%
					\frac{\cosh(\e_+/2T)}{\cosh(|E_+|/2T)}\right]%
							+\ln\left[%
					\frac{\cosh(\e_-/2T)}{\cosh(E_-/2T)}%
				\right]\right\}\nonumber\\
			& &+\frac{1}{2}\int\frac{d\bfm{q}}{(2\pi)^3}%
							\Big\{\frac{\Delta_+^2}{2\e_+}\tanh\left(\frac{\e_+}{2T}\right)%
				 		  +\frac{\Delta_-^2}{2\e_-}\tanh\left(\frac{\e_-}{2T}\right)\Big\}\biggl].\label{eq:CJT2}
\eeq
Roughly speaking, the first term represents the thermodynamic energy gain by 
the dynamical quark loop in the presence of the condensate, while the second term represents 
the energy cost to put the condensate in the Fermi gas. $2(N_c-1)/3N_c$ in front of right hand side
of the above equation indicates that the flavors joining the pairing are 2 out of 3, 
while the active colors are $N_c-1$ out of $N_c$. 

In the CFL case, the algebra is more complicated but a straightforward calculation
yields the following result,
\beq
		\frac{\delta\Omega_{\rm CFL}}{6N_c}%
		&=&-T\int\frac{d\bfm{q}}{(2\pi)^3}%
			\Big\{\frac{1}{N_c^2}%
  				\ln\left[\frac{\cosh(\e_{1+}/2T)}{\cosh(|E_{+}|/2T)}\right]
 			+\frac{N_c^2-1}{N_c^2}%
					\ln\left[\frac{\cosh(\e_{8+}/2T)}{\cosh(|E_{+}|/2T)}\right]%
			\Big\}\nonumber\\
		& &-T\int\frac{d\bfm{q}}{(2\pi)^3}%
			\Big\{\frac{1}{N_c^2}%
  				\ln\left[\frac{\cosh(\e_{1-}/2T)}{\cosh(E_{-}/2T)}\right]
 			+\frac{N_c^2-1}{N_c^2}%
					\ln\left[\frac{\cosh(\e_{8-}/2T)}{\cosh(E_{-}/2T)}\right]%
			\Big\}\nonumber\\
		& &+\frac{1}{2}\int\frac{d\bfm{q}}{(2\pi)^3}%
							\Big\{%
							\frac{1}{N_c^2}\frac{\Delta_{1+}^2}{2\e_{1+}}%
									\tanh\left(\frac{\e_{1+}}{2T}\right)%
				 		  +\frac{N_c^2-1}{N_c^2}\frac{\Delta_{8+}^2}{2\e_{8+}}%
									\tanh\left(\frac{\e_{8+}}{2T}\right)\Big\}\nonumber\\
  		& &+\frac{1}{2}\int\frac{d\bfm{q}}{(2\pi)^3}%
							\Big\{%
							\frac{1}{N_c^2}\frac{\Delta_{1-}^2}{2\e_{1-}}%
									\tanh\left(\frac{\e_{1-}}{2T}\right)%
				 		  +\frac{N_c^2-1}{N_c^2}\frac{\Delta_{8-}^2}{2\e_{8-}}%
									\tanh\left(\frac{\e_{8-}}{2T}\right)\Big\}.\label{eq:CJT3}
\eeq
Eqs.~(\ref{eq:CJT3}) and (\ref{eq:CJT2}) take simple forms
$\delta\Omega_{\rm CFL}(\mu,T)/\mu^4%
=f(\Delta_1/\mu)+(N_cN_f-1)f(\Delta_8/\mu)%
=f(\Delta_1/\mu)+8f(\Delta_8/\mu)$
and 
$\delta\Omega_{\rm 2SC}(\mu,T)/\mu^4=2(N_c-1)f(\Delta/\mu)%
=4f(\Delta/\mu)$. Therefore, if the gaps $\Delta, \Delta_{1}$ and $\Delta_8$
take all the same value, then the condensation energy for the CFL phase
is $9/4$ times larger than that for the 2SC phase.
The comparison of the $(u,d)+s$ 2SC matter with the $(u,d,s)$ 
CFL phase is not quite obvious till one adopts a model and solves
the gap equations and the thermodynamic potential.
\subsection{Occupation numbers and Correlation functions}
Here we give expressions for various quantities in the ground state.

{\it Quark number density~:~~}Occupation 
number characterizes how much the Fermi sphere is distorted by
the gap. We can extract the occupation numbers from the frequency integral 
of the diagonal propagator $S_{11}(q_0,q)$,
\beq
		&&n_+^{r,b}(q)=\frac{1}{2}-\frac{E_+}{2\e_+}\tanh\left(\frac{\e_+}{2T}\right),%
				\,\,n_-^{r,b}(q)=\frac{1}{2}-\frac{E_-}{2\e_-}\tanh\left(\frac{\e_-}{2T}\right)\\
		&&n_+^g(q)=(1-n_F(|E_+|)\theta(-E_+)+n_F(|E_+|)\theta(E_+),\,\,n_-^g(q)=n_F(E_-).\label{baka}
\eeq
$n_+(q)$ and $n_-(q)$ are distribution functions for quark and antiquark.
$n_F(|E_+|)\theta(-E_+)$, $n_F(|E_+|)\theta(E_+)$ and $n_F(E_-)$
represent the thermal hole, the thermal quark and the antiquark, respectively.
Net quark density for the 2SC in some chemical potential $\mu$ can be obtained by
\beq
	\rho(\mu)&=&4\int\frac{d\bfm{q}}{(2\pi)^3}%
		\Big[(N_c-1)(n_{+}^{r,b}(q)-n_-^{r,b}(q))+(n_{+}^{g}(q)-n_-^g(q))\Big].
\eeq
The chemical potential has to be determined by the number conservation
$\rho=\rho(\mu)$, which also can be obtained by the derivative of the 
pressure $-\Omega(\mu)$.

Similarly, we can find occupation number in the CFL phase as
\beq
	n_{1\pm}(q)=\frac{1}{2}-\frac{E_{1\pm}}{2\e_{1\pm}}\tanh\left(\frac{\e_{1\pm}}{2T}\right)%
	,\,n_{8\pm}(q)=\frac{1}{2}-\frac{E_{8\pm}}{2\e_{8\pm}}\tanh\left(\frac{\e_{8\pm}}{2T}\right).
\eeq
$n_1$ and $n_8$ are the color-flavor singlet and octet distribution, which
can be defined using the annihilation operator for quark, $a(p)$ and 
that for antiquark $b(p)$ as
\beq
	\langle a_{i}^{a\dagger}(p) a_j^b(p)\rangle&=&(P_1)^{ab}_{ij}n_{1+}(p)+(P_8)^{ab}_{ij}n_{8+}(p),\nonumber\\
	\langle b_{i}^{a\dagger}(p) b_j^b(p)\rangle&=&(P_1)^{ab}_{ij}n_{1-}(p)+(P_8)^{ab}_{ij}n_{8-}(p).\nonumber
\eeq
Using these octet and singlet densities, we find by putting $b=a$ and $j=i$
in the above equation,
\beq
	n^{a}_{i\pm}(q)=n_{8\pm}(q)+\delta^a_i\frac{1}{N_c}\Big[{n_{1\pm}(q)-n_{8\pm}(q)}\Big].
\eeq
$n^a_i(p)$ is the probability distribution finding a quark $(+)$ or
an antiquark $(-)$ with color $a$ and flavor $i$ having the momentum $p$.
We can see the fact that the quark density is almost determined by the octet density.
The total quark numbers for quark and antiquark read
\beq
	\rho(\mu)&=&2\int\frac{d\bfm{q}}{(2\pi)^3}%
	\Big[(N_cN_F-1)(n_{8+}(q)-n_{8-}(q))+(n_{1+}(q)-n_{1-}(q))\Big].
\eeq

{\it Correlation functions~:~~}What 
is important to characterize a superconducting phase is a energy gap 
and the correlation functions which describes the off-diagonal long 
range order in the system. This quantities are also called the anomalous density
because it is related to the anomalous propagators $\langle Tqq\rangle$.
In the case of the 2SC, 
\beq
	\phi_{\pm}(q)=\tanh\left(\frac{\e_\pm}{2T}\right)\frac{\Delta_\pm}{2\e_\pm}%
			=\Big[1-2n_F(\e_\pm)\Big]\frac{\Delta_\pm}{2\e_\pm}.\label{eq:lkj}
\eeq
Here, the correlation functions $\phi_\pm(p)$ are defined by
\beq
		\langle a^a_{i}(p)a^b_{j}(-p)\rangle=(\tau_2\lambda_2)^{ab}_{ij}\phi_+(q),\,\,%
		\langle b^a_{i}(p)b^b_{j}(-p)\rangle=(\tau_2\lambda_2)^{ab}_{ij}\phi_-^*(q).\nonumber
\eeq
To avoid an involved notation, we have omitted the helicity indices.
As is postulated for the ansatz of the 2SC condensate, 
the correlation exists only in the color-flavor anti-symmetric channel.
The minus sign in front of $n_F$ in Eq.~(\ref{eq:lkj}) shows the effect that 
the thermally excited quasi-particles reduce the strength of the correlation.

In the same way as in the 2SC case, we easily derive the correlation functions 
	in the CFL phase,
The anomalous densities in the singlet and octet channels become
\beq
	\phi_{1\pm}(q)=\tanh\left(\frac{\e_{1\pm}}{2T}\right)\frac{\Delta_{1\pm}}{2\e_{1\pm}},%
	\,\,\phi_{8\pm}(q)=\tanh\left(\frac{\e_{8\pm}}{2T}\right)\frac{\Delta_{8\pm}}{2\e_{8\pm}}.
\eeq
with definition
\beq
		\langle a^a_i(p)a^b_j(-p)\rangle&=&(P_1)^{ab}_{ij}\phi_{1+}(q)+(\bar{P}_8)^{ab}_{ij}\phi_{8+}(q),\nonumber\\%
		\langle b^a_i(p)b^b_j(-p)\rangle&=&(P_1)^{ab}_{ij}\phi_{1-}^*(q)+(\bar{P}_8)^{ab}_{ij}\phi_{8+}^*(q).\nonumber
\eeq
By using the locking relation 
$\e_{Iab}\e^{I}_{ij}=(N_c-1)(P_1)^{ab}_{ij}-(\bar{P}_8)^{ab}_{ij}$,
the color $\bar{\bf 3}_c$ flavor antisymmetric correlation $\phi_A(q)$ and 
the color $\bfm{6}_c$ and flavor symmetric correlation $\phi_S(q)$ are extracted as
\beq
	\phi_A(q)=\frac{1}{N_c}\phi_{1\pm}(q)-\left(1+\frac{1}{N_c}\right)\phi_{8\pm}(q),\,\,%
		\phi_S(q)=\frac{1}{N_c}\phi_{1\pm}(q)+\left(1-\frac{1}{N_c}\right)\phi_{8\pm}(q).\nonumber
\eeq
We would see $\phi_S$ disappears rapidly near the critical temperature
compared with $\phi_A$, due to anti-triplet dominance in the critical region.

\section{Numerical results}\label{sec:3}
In this section, we present our numerical results.
First, we focus on how the temperature affects the
superconducting gap, and other physical quantities at
		a high chemical potential $\mu=1000$~MeV.
We discuss various features of the phase transition.
Then, we see how these pictures of the phase transition 
would be modified towards low, physically interesting densities. 
Finally, we draw the QCD phase diagram for the region $\mu\agt400$~MeV.
We estimate the effect of the effective strange quark mass $m_s$ on the QCD 
phase diagram using a simple unlocking criterion, and study how the
QCD phase diagram changes with a variation of $m_s$. 

\subsection{Phase transitions to QGP phase at $\mu=1000$~MeV}
\begin{figure}[tbp]
 \vspace{2mm}
 \begin{minipage}{0.49\textwidth}
  \epsfsize=0.99\textwidth
  \epsfbox{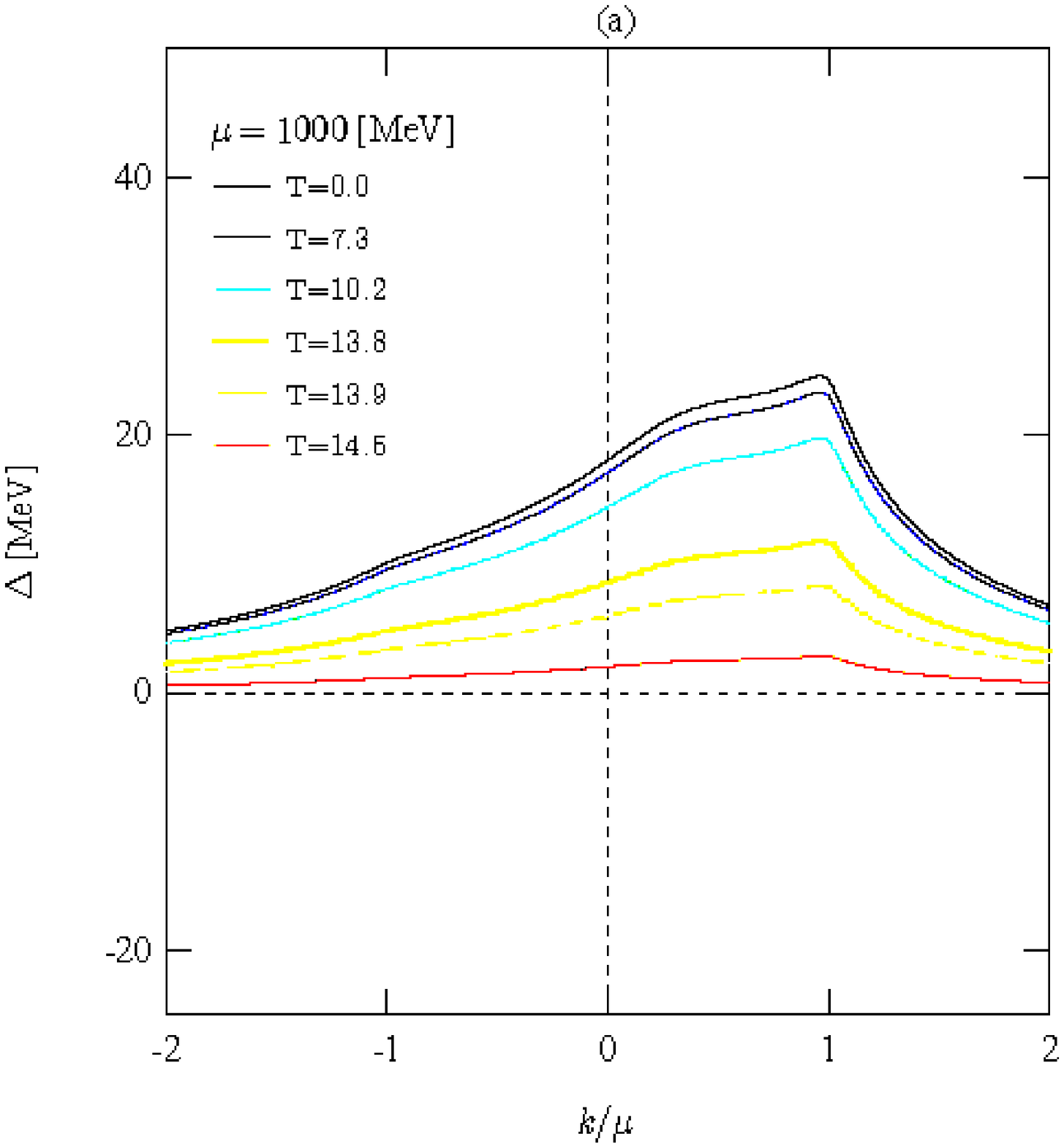}
 \end{minipage}%
 \hfill%
 \begin{minipage}{0.49\textwidth}
  \epsfsize=0.99\textwidth
  \epsfbox{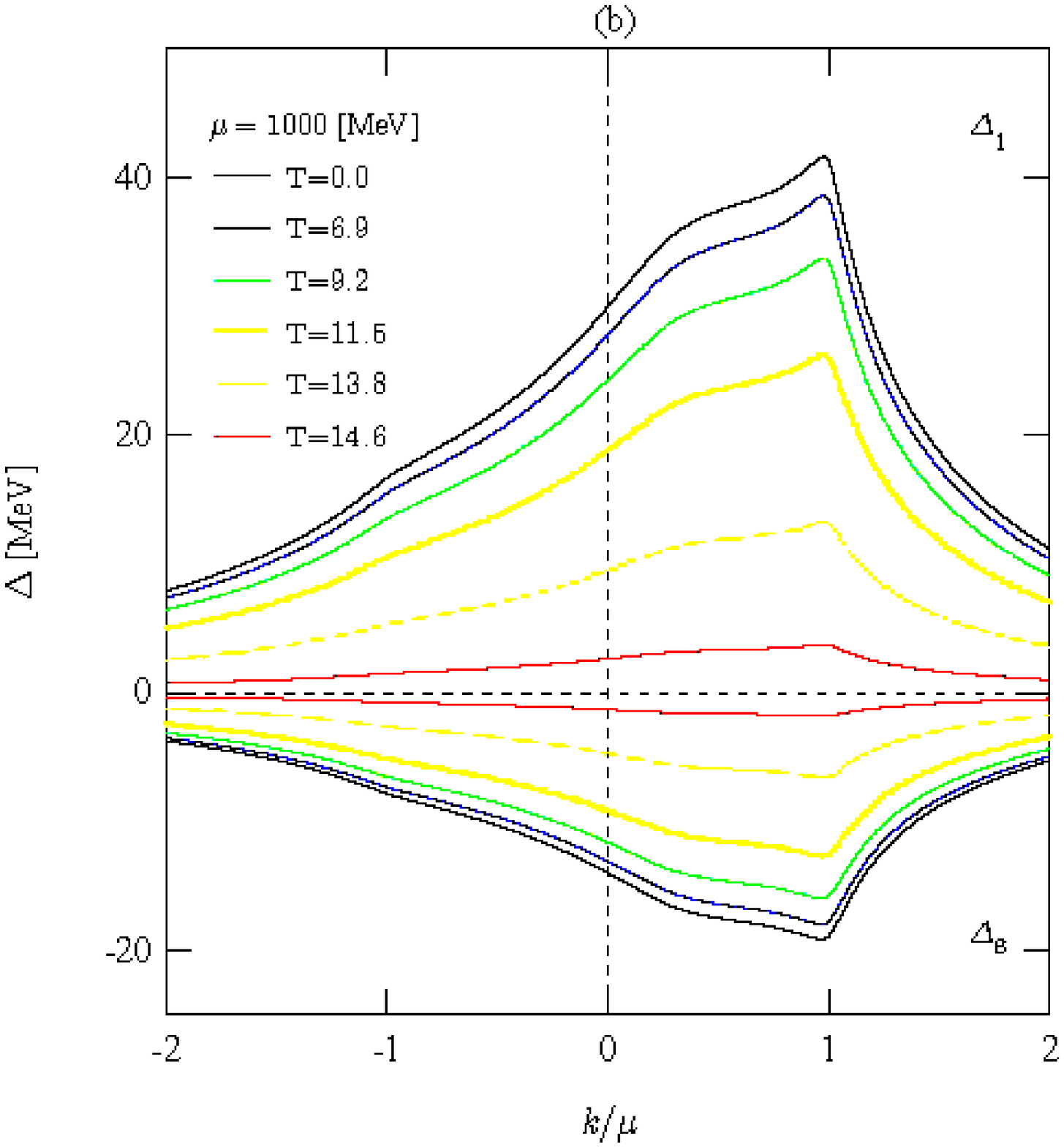}
 \end{minipage}%

 \caption[]{Gap functions at $\mu=1000$~MeV and various temperatures.
						We have drawn the gap for antiquarks at negative momentum
						region, namely, $(-|k|, \Delta_-(|k|))$ are plotted at left
						half of the figure. 
					(a) Gap function for the 2SC state.
					(b) Gap functions for the CFL state.}
 \label{fig:1}
\end{figure}

{\it Gap functions and condensation energies~:~~}%
We show the numerical solutions for the gap equations
	Eqs.~(\ref{eq:001})$\sim$(\ref{eq:002}) at a fixed chemical 
	potential $\mu=1000$~MeV corresponding to about 80 times
	the normal nuclear density $\rho_0=0.17~{\rm fm}^{-3}$.
Fig.~\ref{fig:1} shows the gap function $\Delta_\pm(p)$ for 
	the 2SC state (a), and the singlet and octet gap functions 
	$(\Delta_{1\pm}(p),\Delta_{8\pm}(p))$ for the CFL state (b).
At first sight, we can see the shapes of these gap functions 
	are all similar as a function of momentum except for the
	magnitude $|\Delta_8(p)|<|\Delta(p)|<|\Delta_1(p)|$ and the 
	relative sign between the singlet and octet gaps.
Secondly, we observe the temperature uniformly reduces the
	magnitudes of the gap functions towards the critical 
	temperature which can be read $\sim 15$~MeV. 

\begin{figure}[tbp]
 \vspace{2mm}
 \begin{minipage}{0.49\textwidth}
  \epsfsize=0.99\textwidth
  \epsfbox{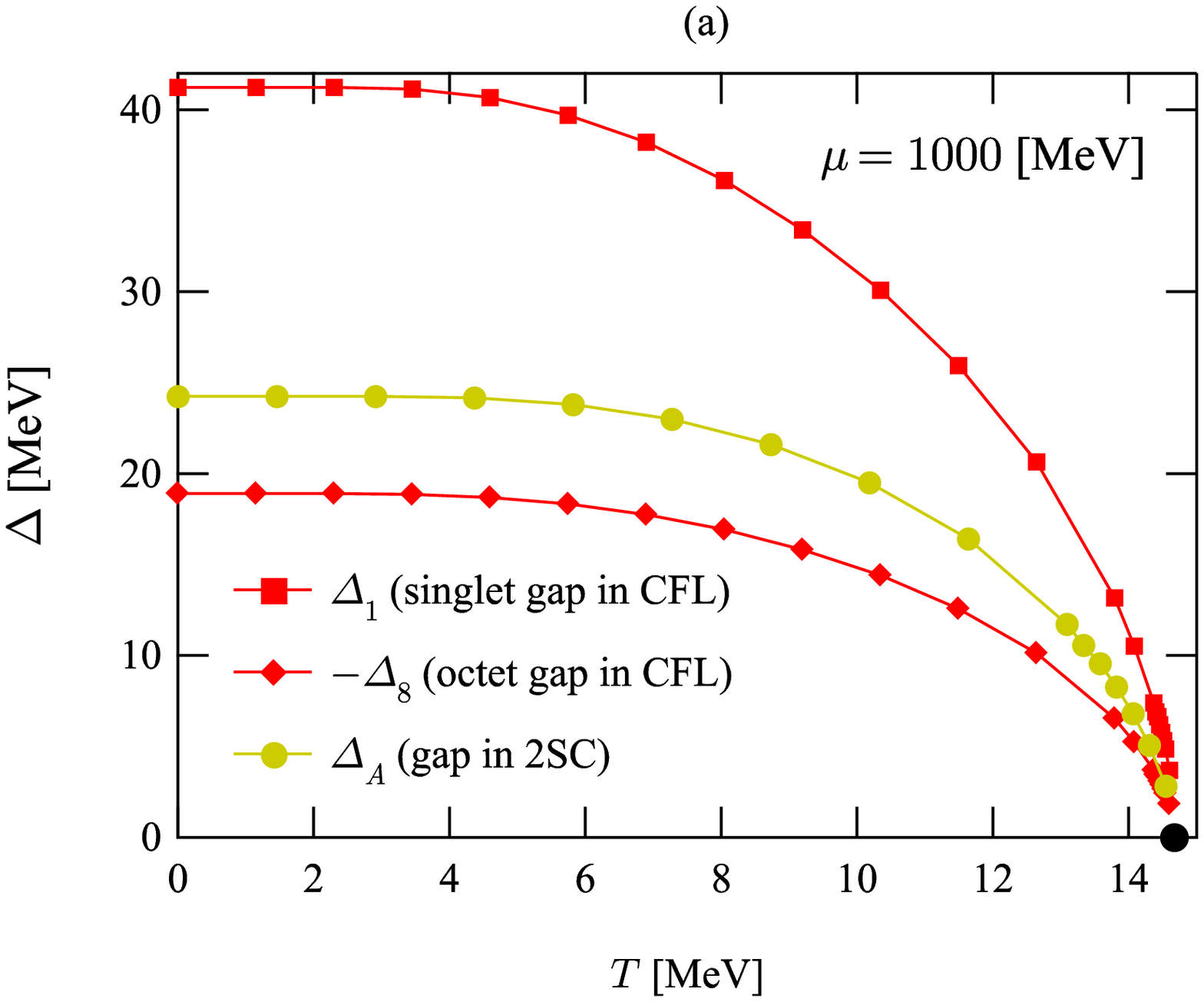}
 \end{minipage}%
 \hfill%
 \begin{minipage}{0.49\textwidth}
  \epsfsize=0.99\textwidth
  \epsfbox{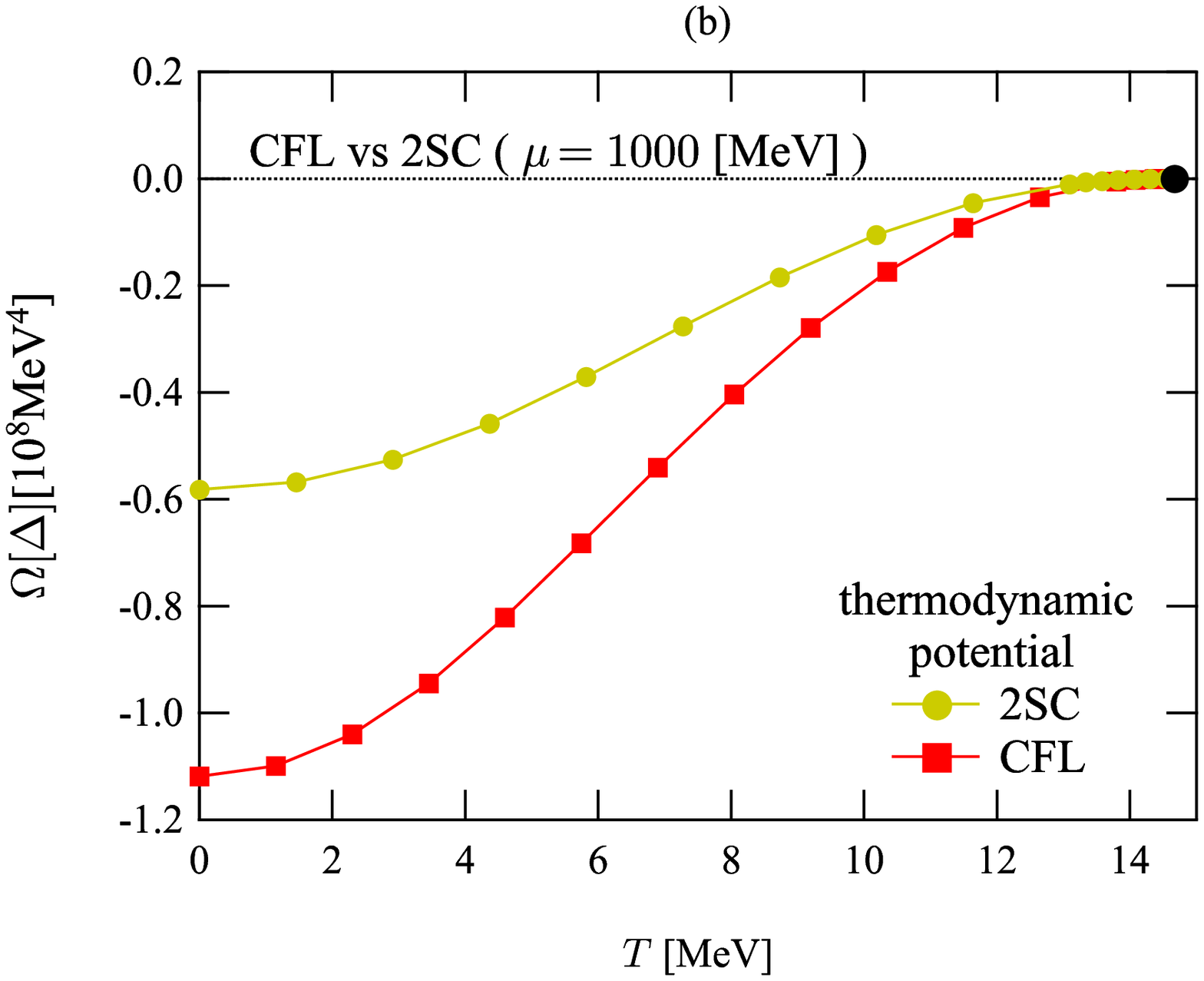}
 \end{minipage}%
 \caption[]{CFL vs. 2SC.

					(a) Gaps at the Fermi surface vs. temperature.
					(b) Thermodynamic potential and energy gain per quark
							for both phases at $\mu=1000$~MeV.
}
 \label{fig:2}
\end{figure}

In order to see characteristics of phase transitions in more detail,
	we show in Fig.~\ref{fig:2}, the temperature dependence of the gaps 
	at Fermi surface (a) and that of the CJT condensation energies for the 
	2SC and CFL states (b).
From these figures, we can see that the phase transitions to the QGP phase
	both from the 2SC state and from the CFL phase are of 2nd order just like 
	in the BCS theory with contact interaction : 
	The derivatives of the gaps seem to diverge towards the $T_c$ in (a), 
	and thermodynamic bulk quantities in superconducting
	phases are connecting continuously to those of normal QGP phase (b).
Another thing to be noticed is that the critical temperatures
	to the QGP phase are the same both for the CFL and for the 2SC.
We will discuss this property in more detail latter in this section
	and see that this holds at least in the weak coupling limit.

The condensation energy for the CFL state is always about 1.9$\sim 2$ times larger than 
	that for the 2SC in spite of the fact that the octet gaps determining the thermodynamics 
	in the CFL takes smaller values than the 2SC gaps.
This is consistent with the weak coupling analytical estimate \cite{Schafer1999}~:~%
The CFL phase is easier to excite quasi-quarks, but the bulk energy is lower in the CFL 
	than in the 2SC.
This reversal between the condensation energy and the magnitude of 
	gap is clearly a direct consequence of the fact that the number of the
	active freedom taking part in pairing is $9/4$ times larger in the
	CFL phase than in the 2SC phase\footnote{The condensation energy for the CFL is a function of $\Delta_1^2+8\Delta_8^2\sim 12\Delta_8^2$, 
	and that for the 2SC is a function of $4\Delta^2$. Thus, 
	the thermodynamic enhancement factor $9/4$ is a little bit underestimated, 
	but is enough for explaining the CFL dominance at $T=0$.}
It can be said that the CFL state is relatively more fragile.\footnote{Because the quasi-quarks 
	contribution to the thermodynamic quantities in the CFL can be expressed 
	as the integral of $8e^{-\e_8/T}+e^{-\e_1/T}$, the color-flavor singlet quarks 
	with the large gap do not give a significant thermal effect for $T\alt T_c$.
	On the other hand, in the 2SC state, thermal quark contribution to the
	thermodynamic quantities can be written as the integral of $4e^{-\e/T}$.
	Thus, the CFL state is more sensitive to temperature than the 2SC state 
	because of octet quasi-quarks with the smallest gap.} 
	than the 2SC state due to smaller gap in the color-flavor octet channel.
Reflecting this, the critical temperatures to the QGP phase from both states are almost the same 
	although the condensation energy for the CFL at $T=0$ is $1.9$ times larger than that for the 2SC state.

\begin{figure}[tbp]
 \vspace{2mm}
 \begin{minipage}{0.49\textwidth}
  \epsfsize=0.99\textwidth
  \epsfbox{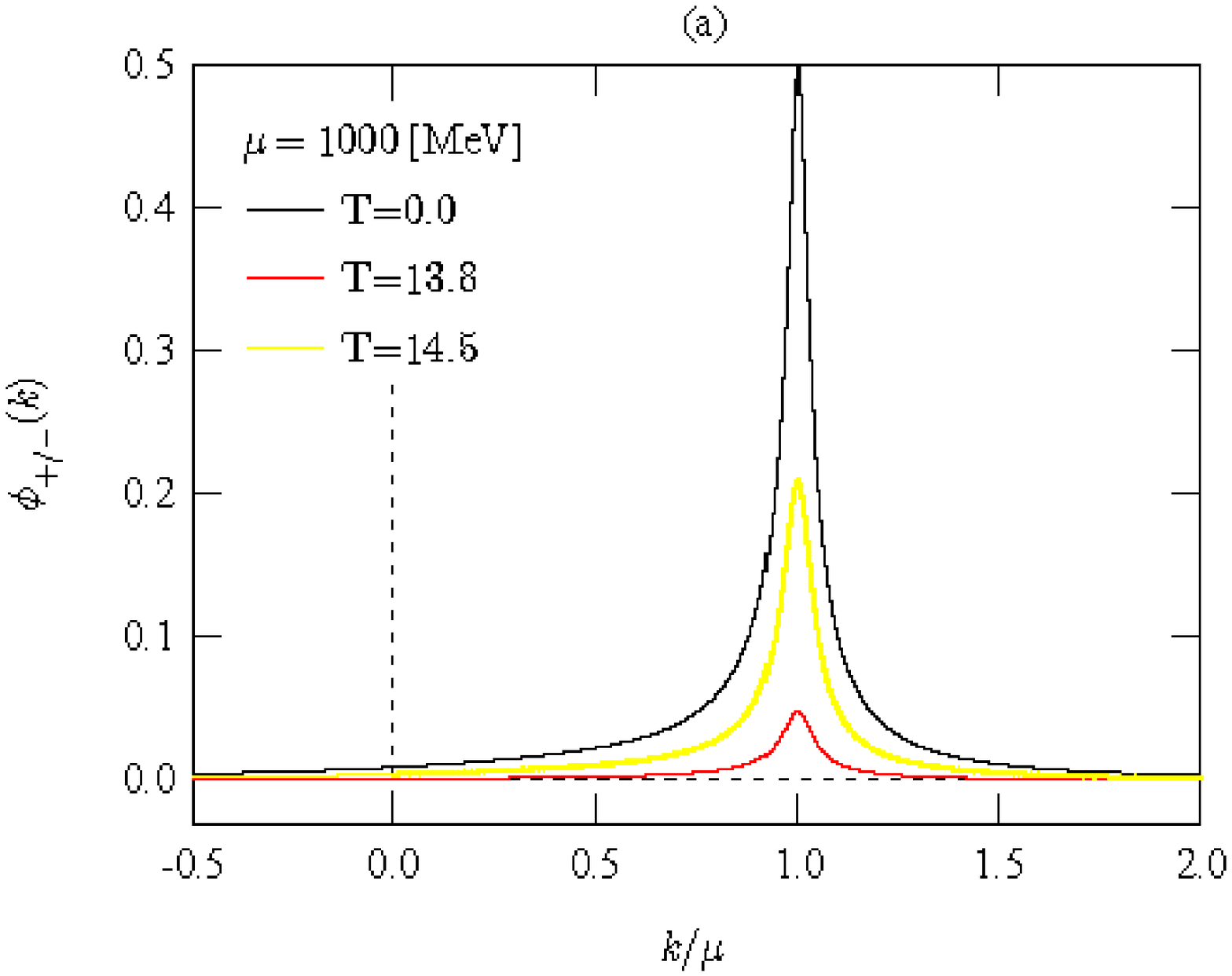}
 \end{minipage}%
 \hfill%
 \begin{minipage}{0.49\textwidth}
  \epsfsize=0.99\textwidth
  \epsfbox{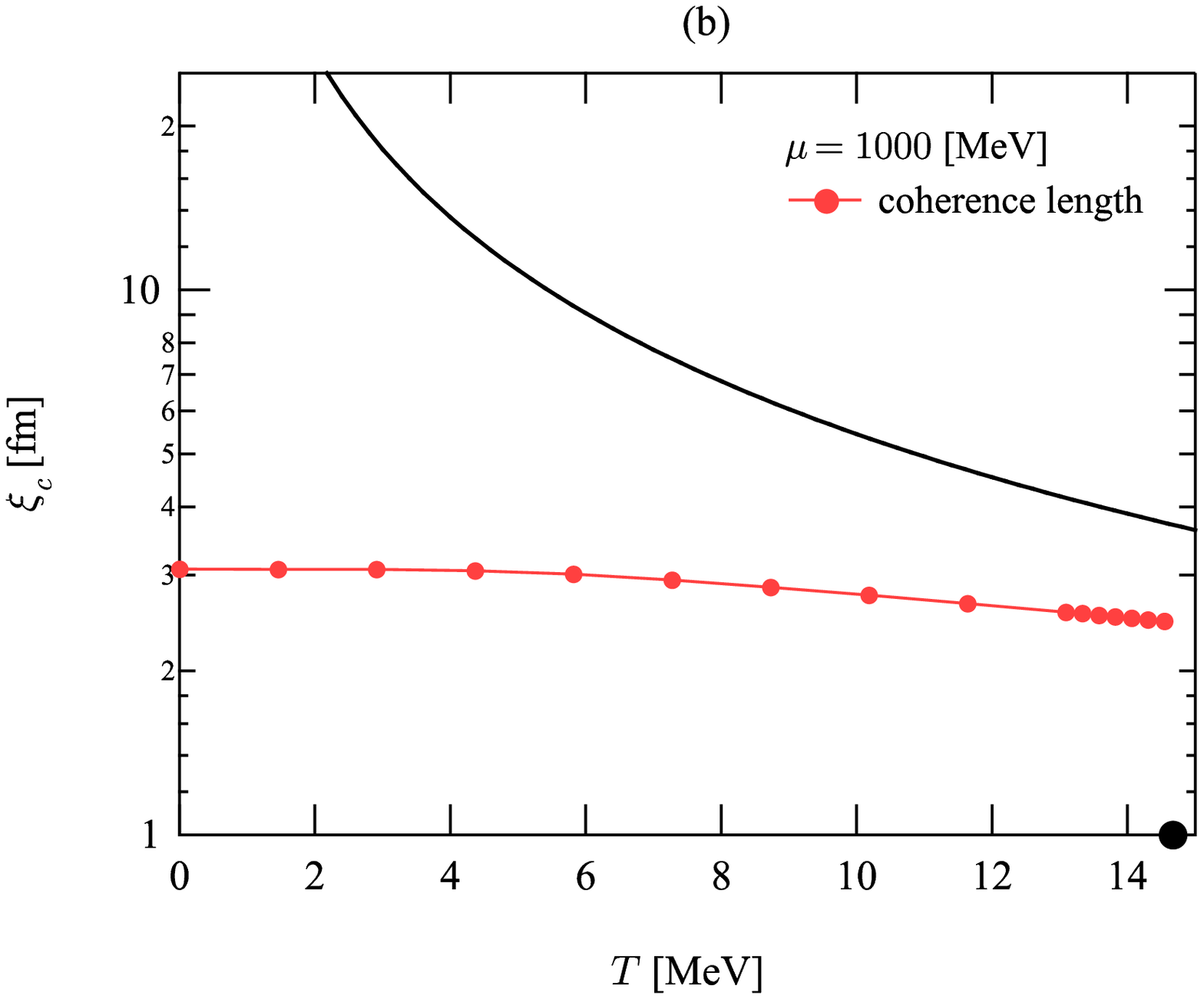}
 \end{minipage}%
 \caption[]{(a) Correlation functions for the 2SC at finite temperature.
					(b) The coherence length and the Pippard length in the 2SC state.
							Black line show the thermal wave length of the un-perturbed 
							Fermi gas $\sim1/2T$.}
 \label{fig:3}
\end{figure}

{\it Cooper pair size at $\mu=1000$~MeV~:~~}In 
Fig.~\ref{fig:3}(a), the correlation functions for the 2SC
	at various temperature are shown.
These quantities characterize the internal structure of pairs.
Although the magnitude of the correlation gets significantly
	reduced by temperature, the width of the function seems not
	so much affected.
This implies that thermally excited quasi-qaurks destroy 
	the phase coherence of the system, but does not change the
	internal structure of pairs.
This fact gives us a picture of the phase transition~:~The 
	phase transition occurs by the reduction of the 
	number of coherent pairs, not by the dissociation of them.
In order to make sure of this point, we have drawn various length
	and those dependence on temperature in Fig.~\ref{fig:3}(b).
As is speculated above, the coherence length $\xi_c$ denoted by black points
	are not much affected by temperatures. $\xi_c$ is defined by
the root mean square radius of the Cooper pair with wave function $\phi_+$.
Black line shows $1/2T$, which is related to the thermal de Broglie 
	wavelength of the naive Fermi gas 
	: $\lambda_T=1/\sqrt{\langle E_+^2\rangle_T}\sim 1/2T$.
Only the interaction with the length scale $\alt\lambda_T$ can cause quantum 
	corrections to the naive Fermi gas.
Phase transition to the QGP phase takes place at these 2 length
	scale becomes comparable, because the quantum correlation with length $\xi_c$ 
	gets gradually destroyed by thermal disturbance with length $\lambda_T$.

\begin{figure}[tbp]

 \vspace{2mm}
 \begin{minipage}{0.49\textwidth}
  \epsfsize=0.99\textwidth
  \epsfbox{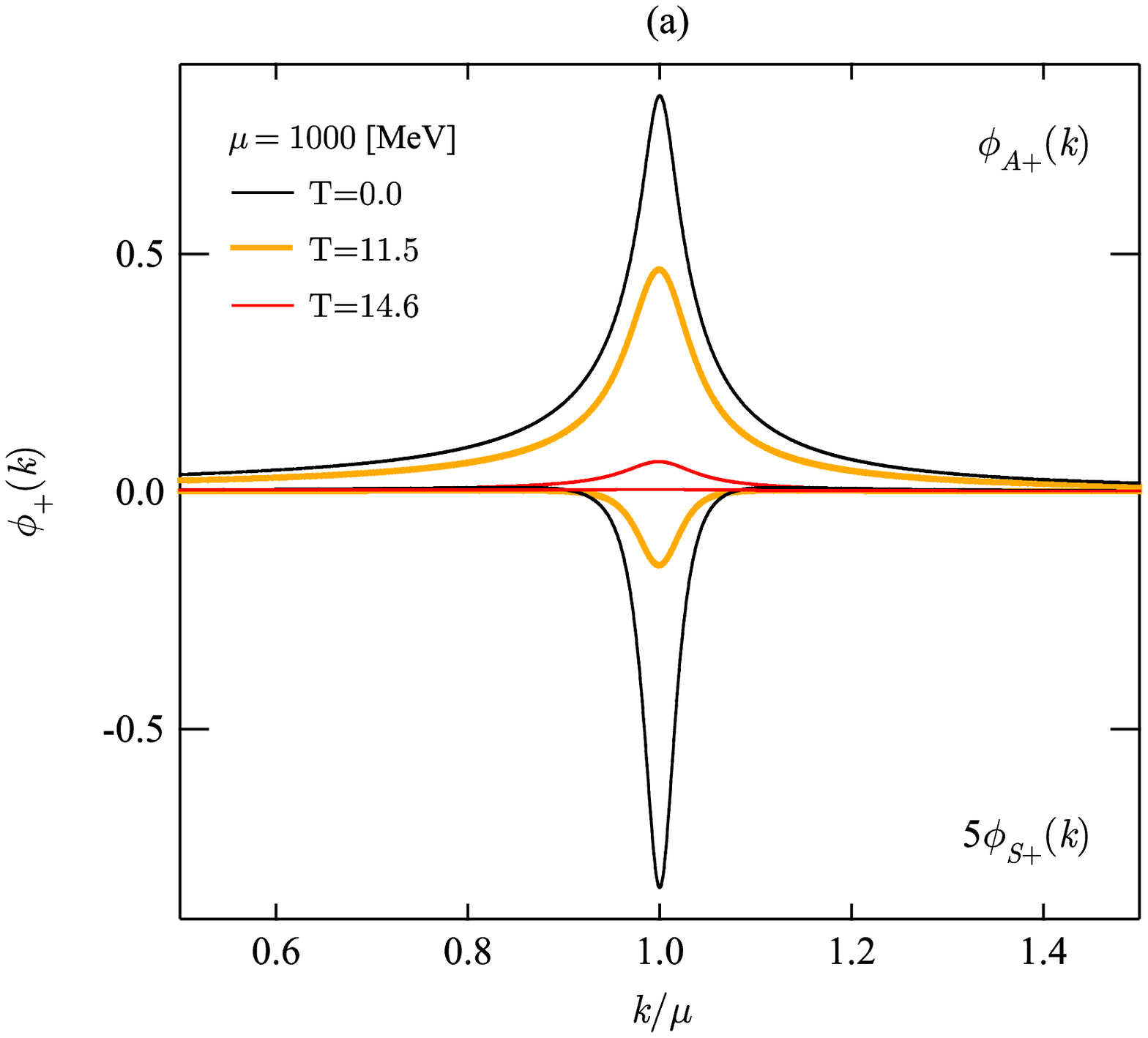}
 \end{minipage}%
 \hfill%
 \begin{minipage}{0.49\textwidth}
  \epsfsize=0.99\textwidth
  \epsfbox{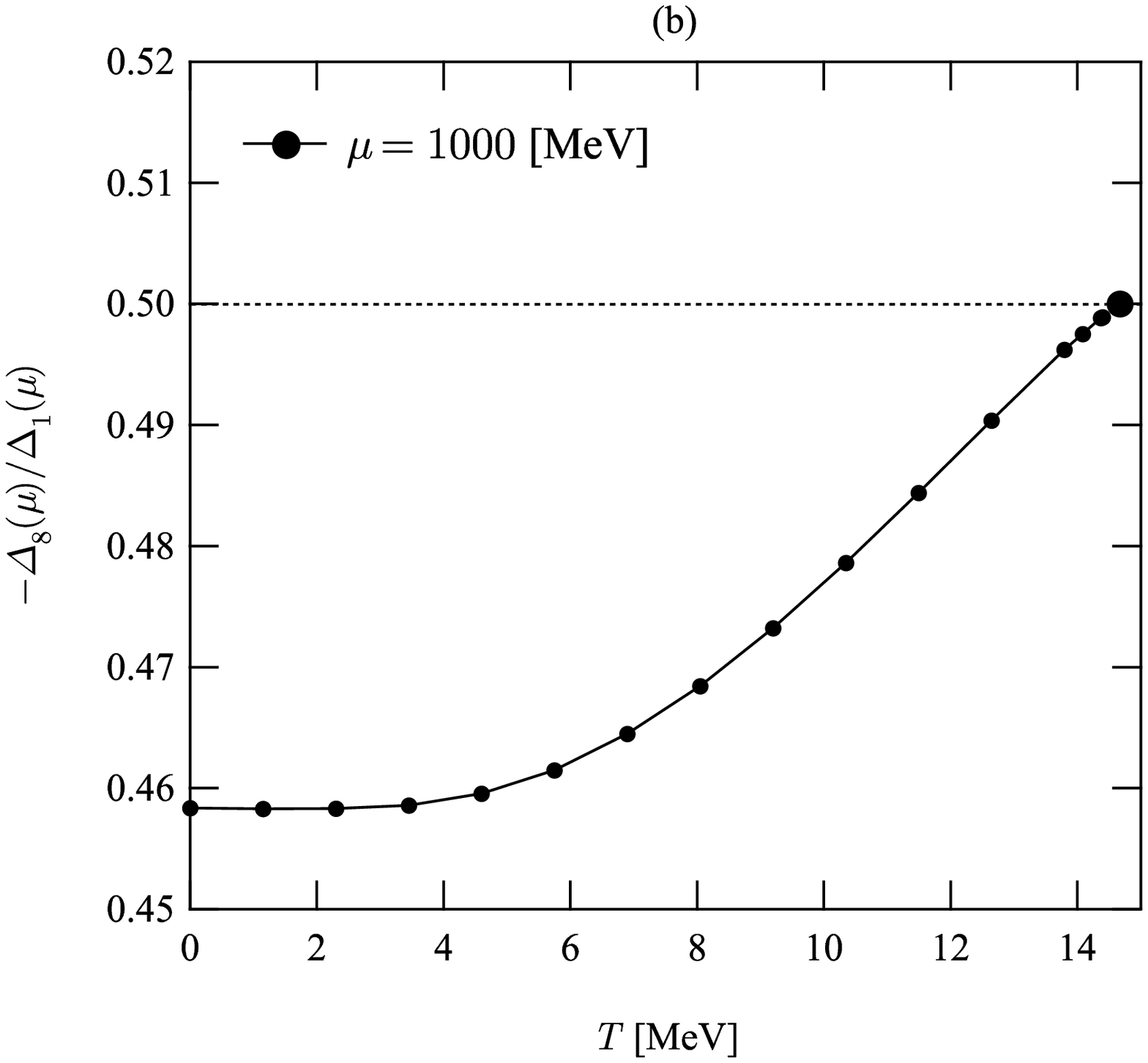}
 \end{minipage}%
 \caption[]{Disappearance of nonlinearity near critical temperature.
					(a) The temperature dependence of correlation functions 
							in the anti-triplet and sextet channels for the CFL state.
					(b) The temperature dependence of the ratio $-\Delta_8/\Delta_1$.}
 \label{fig:4}
\end{figure}

{\it Why critical temperatures coincides?~:~~}Now, 
we return to the problem why the critical temperatures
	to the QGP phase coincides.
It is related to the disappearance of nonlinear effects near critical
	temperature $T_c$.
To see this, we rewrite the CFL gap Eqs.~(\ref{eq:003})$\sim$(\ref{eq:002})
	in terms of the symmetric and anti-symmetric self-energy 
	($\Delta_A$, $\Delta_S$) in Eq.~(\ref{eq:dec2}).
If we neglect the $\Delta_S$ term in the loop integral in Eqs.~(\ref{eq:003}) and (\ref{eq:002})
	(ansatz of the anti-triplet dominance $|\Delta_S|\ll|\Delta_A|$), and also the contribution from antiquarks 
	which can be justified
	only in the weak coupling region, equations for $\Delta_A$ and $\Delta_S$ cast into the following form:
\beq
		\frac{4N^2}{N+1}\Delta_A&=&(N+1)K^+\left[\Delta_A\right]%
															+K^+\left[(N-1)\Delta_A\right],\\
		\frac{4N^2}{N-1}\Delta_S&=&(N-1)K^+\left[\Delta_A\right]%
															-K^+\left[(N-1)\Delta_A\right].
\eeq
If we neglect the nonlinear terms in $\Delta_A$, we obtain
\beq
		\frac{4N^2}{N+1}\Delta_A=2NK^+\left[\Delta_A\right],\quad%
		\frac{4N^2}{N-1}\Delta_S=0.\label{eq:critical}
\eeq
These two ``{\it gap equations}'' are consistent with the first ansatz of omitting 
	$\Delta_S$ in the loop integral. 
Discarding nonlinear terms in $\Delta_A$ is justified near 
	the critical temperature of the 2nd order transition. 
To make sure of the validity of this argument, we check the ansatz of 
the color anti-triplet dominance near the critical temperature. 
Fig.~\ref{fig:4}(a) shows correlation functions in the 
anti-triplet and sextet channels. 
The sextet correlation disappears rapidly towards the critical 
point compared to the anti-triplet correlation.
Similarly, Fig.~\ref{fig:4}(b) shows the temperature dependence
of the ratio $-\Delta_8/\Delta_1$. This quantity approaches
$1/2$ near the critical temperature, which directly means
the anti-triplet dominance $\Delta_S/\Delta_A\to 0$ with $T\to T_c$.
The convergence $-\Delta_8/\Delta_1\to 1/2$ seems linear 
in $t=(T_c-T)/T_c$, therefore, 
if $\Delta_A(t)\cong-\Delta_8(t)\sim t^{1/2}$ near $T_c$
		just as the scaling in the mean field approximation, then $\Delta_S$
		approaches $0$ as $\Delta_S\sim t^{3/2}$ towards $T_c$.
Namely, $\Delta_S$ disappears rapidly towards $T_c$.

\subsection{Gaps, condensation energies and critical temperatures for $\mu=1000\sim 400$~MeV}
Now we turn to the axis of chemical potential in the phase diagram, and investigate how the physical 
		quantities studied above change, and how the picture of phase transition
		gets modified as one goes to lower density ($\mu<1000$~MeV) region. 

\begin{figure}[tbp]
 \vspace{2mm}
 \begin{minipage}{0.49\textwidth}
  \epsfsize=0.99\textwidth
  \epsfbox{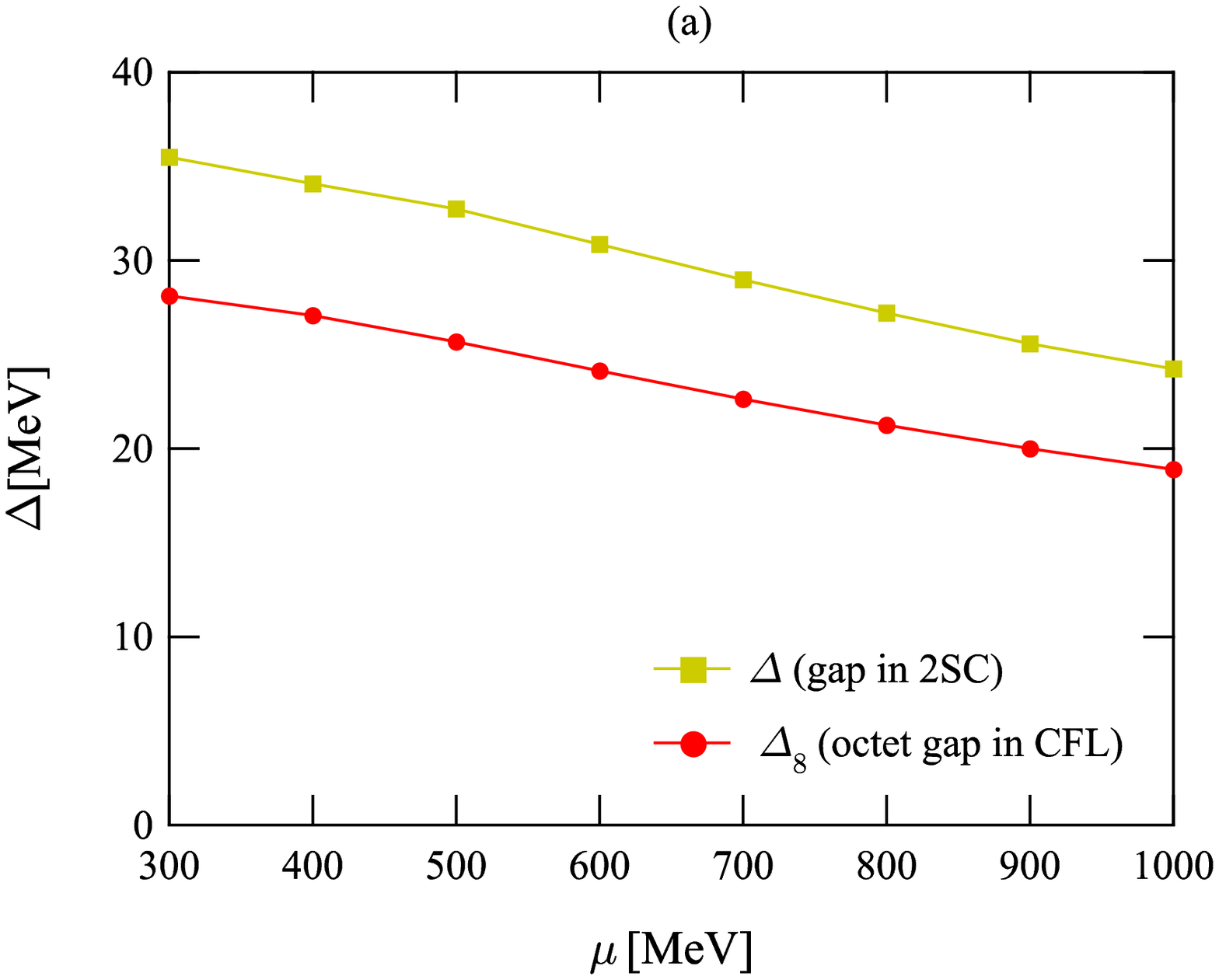}
 \end{minipage}%
 \hfill%
 \begin{minipage}{0.49\textwidth}
  \epsfsize=0.99\textwidth
  \epsfbox{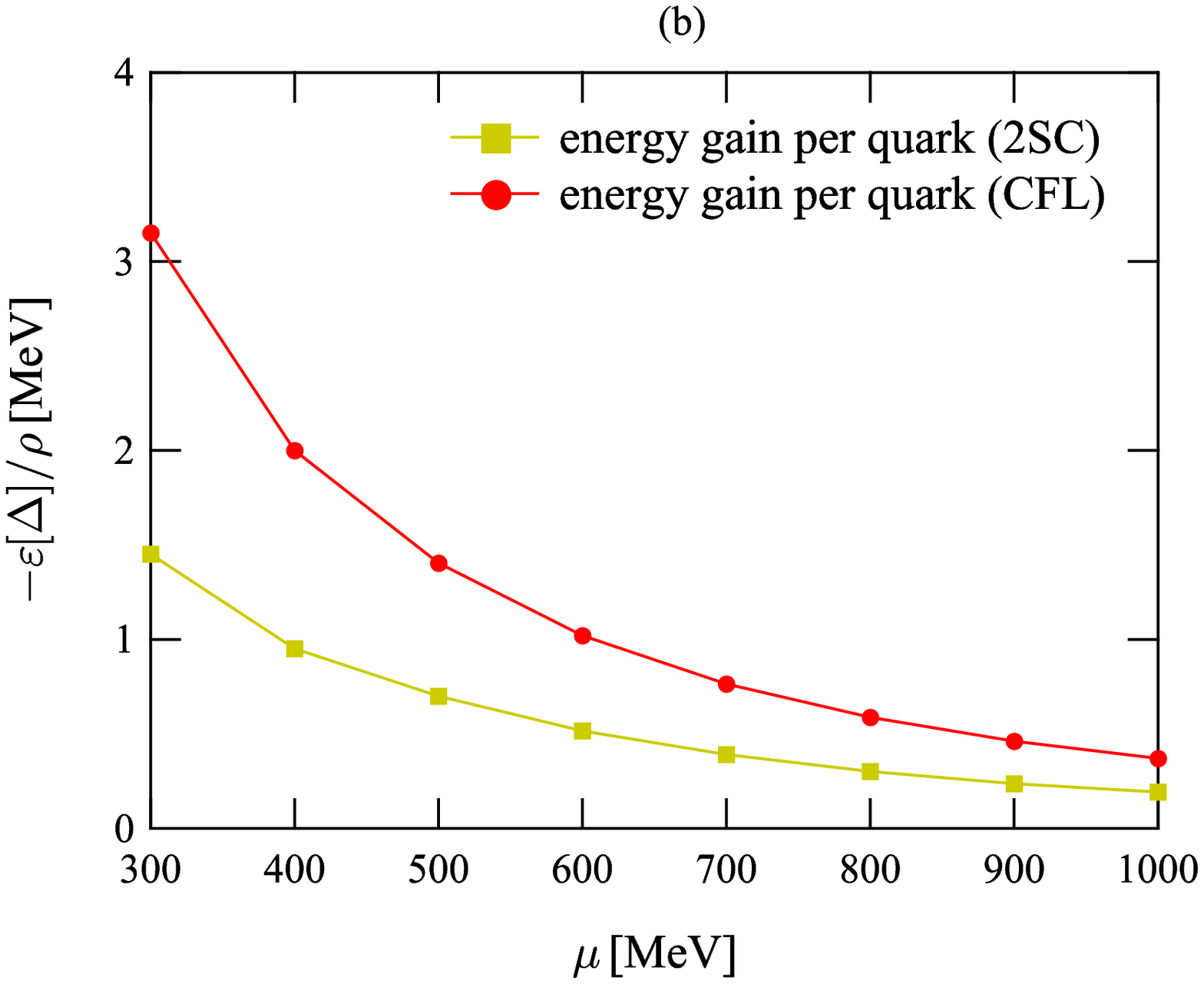}
 \end{minipage}%
 \caption[]{2SC vs. CFL at various densities.
					(a) Chemical potential dependence of the octet gap and the 2SC gap. 
					(b) Density dependence of condensation energy gain per quark 
							both for the CFL and 2SC states.}
 \label{fig:5}
\end{figure}

{\it Gaps and energy gains per quark at $T=0$~:~~}We 
show the chemical potential dependence of the 2SC gap $\Delta$ and the CFL octet gap 
	$|\Delta_8|$ in Fig.~\ref{fig:5}(a).
The ordering $|\Delta_8|<\Delta$ is true for all region $300$~MeV$<\mu<1000$~MeV. 
{Our gaps have several features which are quite different from those obtained 
	in Ref.~\cite{ARW_98}.

\noindent
(1)~The gaps in Ref.~\cite{ARW_98} disappear for $\mu>800$ MeV.
	Because the NJL model used in the reference is an effective theory
	originally introduced to describe the chiral symmetry breaking in the QCD vacuum, 
	it has a cut-off scale $\Lambda\sim 1$ GeV. Our gaps are also decreasing functions 
	of $\mu$, but do not go to zero even at high density.
	Furthermore, our gaps will turn into increasing functions of $\mu$ at some
	higher chemical potential $mu>1000$ MeV, because our model coincide with 
	other perturbative SD approaches \cite{Son_98,Schaefer,Hong,Pisarski-Rischke}, 
	which all predict increasing gaps in the weak coupling regime.

\noindent
(2)~The gaps obtained in Ref.~\cite{ARW_98} go to zero towards $\mu\to 0$,
	while our gaps do not. Gap equation at $\mu=0$ is the same as the chiral
	gap equation under the replacement the NJL 4-Fermi coupling $K\to K/2$,
	if we neglect the small admixture of the pairing correlation in the color and 
	flavor sextet channels. The coupling $K$ has been tuned to reproduce the
	chiral gap $400$ MeV at zero density. Therefore, the disappearance of the 
	superconducting gap at $\mu=0$ means that $K/2<K_c<K$, where $K_c$
	is the critical coupling for the dynamical generation of the chiral mass gap.
	On the other hand, our gaps increase as $\mu\to 0$ and seem to converge 
	to non-zero finite values. 
	It can be said that our model describes a strong coupling superconductor
	at low densities in the sense that the large effective coupling constant leads 
	non-zero gaps even in the absence of the Fermi surface \cite{itakura}.
	Of course, if we had included the diagonal self-enrgy part which is relevant to 
	the chiral symmetry breaking, then one would encounter the phase transition to 
	chiral broken phase at some critical chemical potential $\mu_c$, and the 
	superconducting gap would disappear at zero density due to the reduction of the 
	state density at Fermi surface owing to the generation of mass gap
	\cite{Berges-Rajagopal,Diakonov,Klevansky}.\footnote{However, there is a model 
	in which the diquark gap persists against the sudden jump of the chiral gap \cite{Kitazawa}.}
}

In Fig.~\ref{fig:5}(b), the chemical potential dependence of the energy gain 
per quark relative to the naive Fermi gas is shown. 
First, the CFL state persists against the 2SC for all density region,
and its energy gain is about 1.9 times larger than that in the 2SC.
Second, this quantity also shows an increasing behaviour as one goes to lower density region. 
Nevertheless, it is small in order of $\sim 1$MeV even at $\mu=300$MeV.
This fact is attributed to the fact that particle energy gain is proportional 
to $\sim\Delta(\Delta/\mu)$, namely, the pairing effect on the energy gain
per quark is suppressed by factor $\Delta/\mu$. This means that only $\Delta/\mu$ of total
quarks near the Fermi surface get the energy gain $-\Delta$.
However, $\Delta/\mu\sim 1/10$ at $\mu=300$~MeV is quite larger than the case of 
the BCS weak coupling superconductor, where typically this ratio takes the value $\sim 1/1000$. 
{It can be said that the color superconductor which might be realized in the 
neutron star core is high $T_c$ one of strong coupling.}

\begin{figure}[tbp]
 \vspace{2mm}
 \begin{minipage}{0.49\textwidth}
  \epsfsize=0.99\textwidth
  \epsfbox{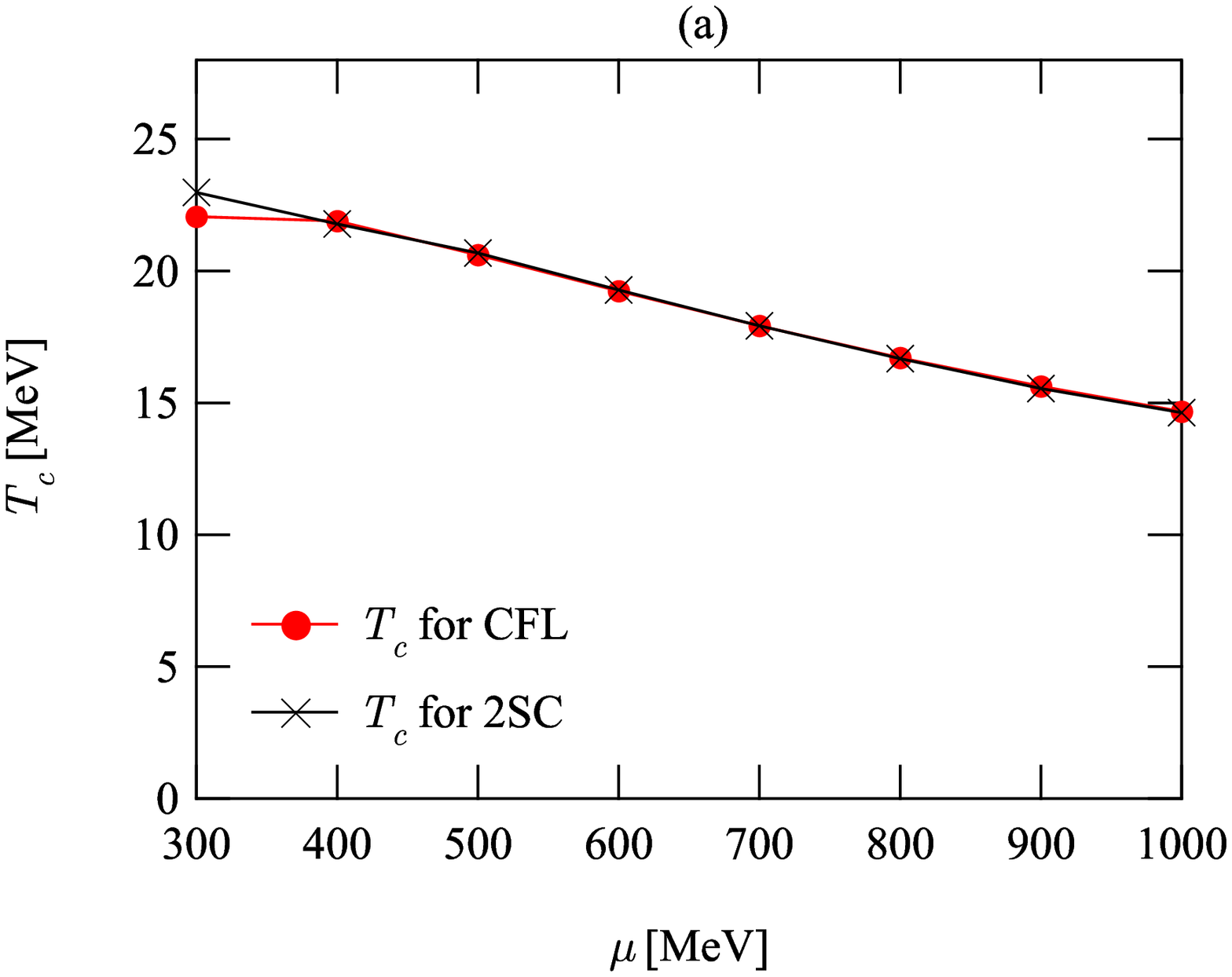}
 \end{minipage}%
 \hfill%
 \begin{minipage}{0.49\textwidth}
  \epsfsize=0.99\textwidth
  \epsfbox{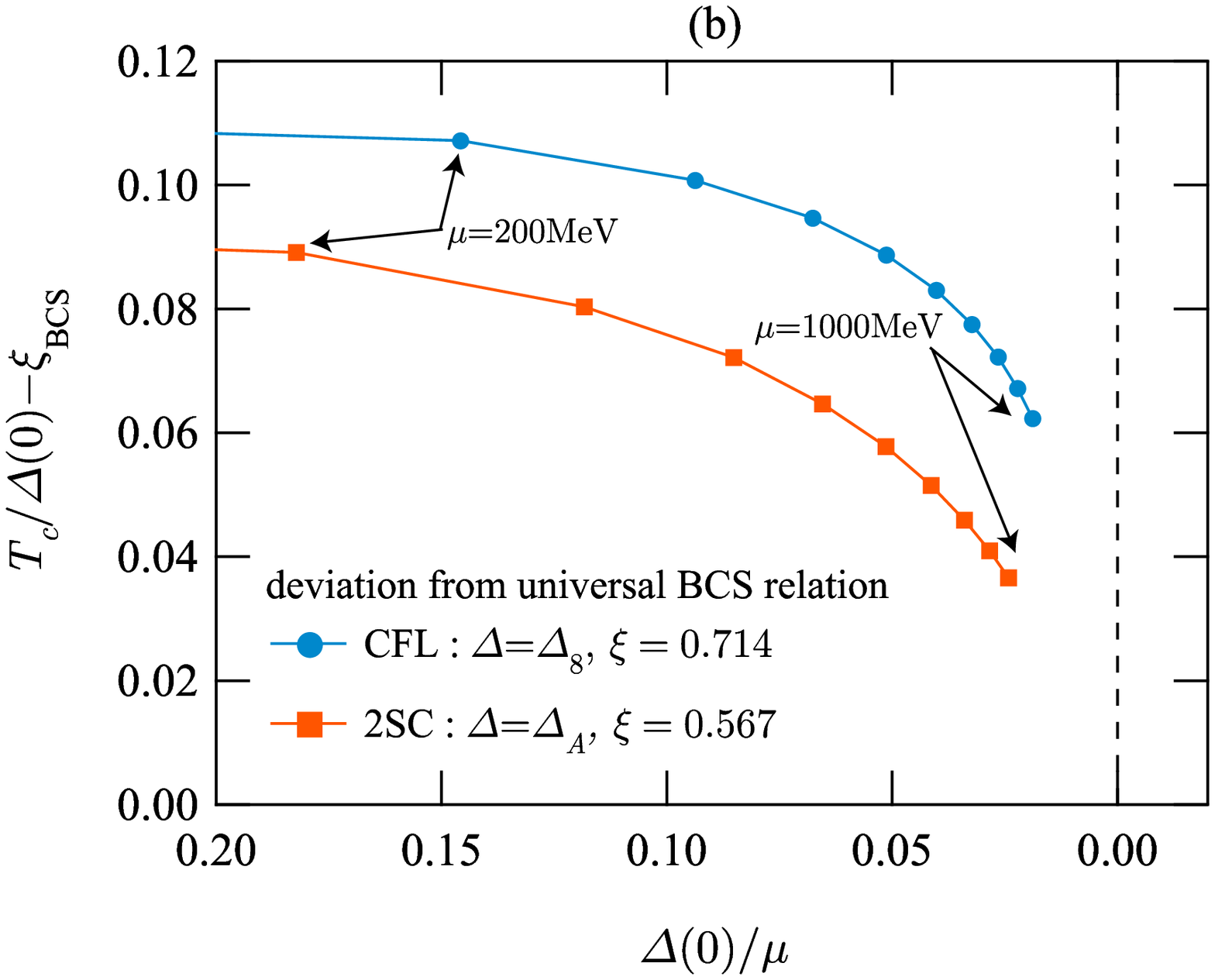}
 \end{minipage}%
 \caption[]{ (a) Critical temperatures extracted by fitting the data in the from 
						$\Delta(T)=\alpha(T_c-T)^\beta$, vs. chemical potential
					(b) The deviations of the ratio $T_c(\mu)/\Delta(\mu,T=0)$ from 
							its weak coupling universal value $\xi$, vs. $\Delta/\mu$, a 
							variable for the weak coupling expansion.
						}
 \label{fig:9}
\end{figure}

{\it Critical temperatures~:~~}Now 
we discuss the chemical potential dependence of the critical temperature.
In Fig.~\ref{fig:9} (a), critical temperatures for the 2SC and CFL states as a
		function of the quark chemical potential are shown.
The dependence of $T_c$ on the chemical potential is similar to those of gaps in Fig.~\ref{fig:5}(a).
The strong coupling effects also exist and lift up critical temperatures at lower densities.
{Even then, the critical temperature at the lowest density is $\sim 20$ MeV.
Although the behaviour of the critical temperature as a function of $\mu$
	is qualitatively similar to that obtained from the improved SD approach
	for $m_s=\infty$ in the Ref.~\cite{Takagi},
the magnitude of the critical temperature is much smaller in our case than
	$\sim 150$ MeV  in the reference.
The correct treatment of the Landau damping in the magnetic gluon would
	make this difference smaller.
}
Fig.~\ref{fig:9}(b) shows the ratio $T_c(\mu)/\Delta(\mu,T=0)$ as a function of $\Delta(\mu,T=0)/\mu$.
This ratio takes the coupling independent universal value in the weak coupling limit $\Delta/\mu\to 0$ 
		in the BCS model with a contact interaction.
We can see that this ratio approaches to the weak coupling value towards $\Delta/\mu\to 0$.
However our interested region is far away from the region where the convergence to the weak coupling 
		values can be seen.
Again, we notice the importance of the strong coupling effects.

Even in the presence of these strong coupling effects at low densities, the ratio of the critical temperatures 
	for the CFL and 2SC states do not deviate from its weak coupling value.
Namely, the critical temperatures for the 2SC state coincides with that for the CFL state 
	for all region of chemical potential within numerical errors. 

{\it Simple estimate of the unlocking transition for $T=0$~:~~}Here 
we estimate the color-flavor unlocking transition at $T=0$ using a simple kinematical criterion, 
which turned out to be quite a good guide for unlocking transition in the NJL-like model analyses \cite{unlock_SW,unlock_AR}.
\begin{figure}[tbp]
  \centerline{
  \epsfsize=0.5\textwidth
  \epsfbox{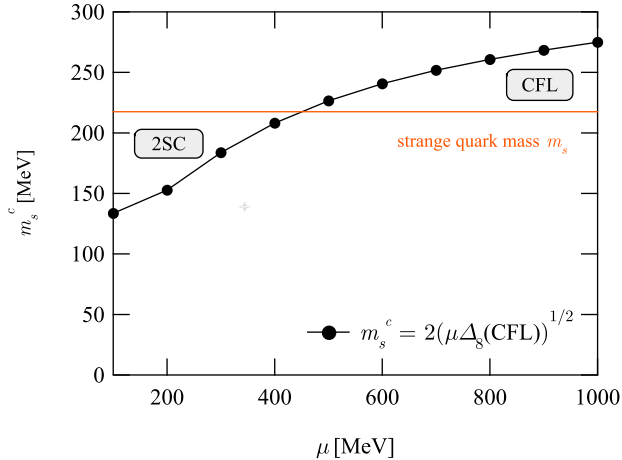}}
 \caption[]%
	{The chemical potential dependence of the critical strange quark mass 
		from a simple criterion for the unlocking transition with 
		$\Delta_8(\mu)$ obtained by solving the CFL gap equations.}
\label{fig:8}
\end{figure}
In Fig.~\ref{fig:8}, we have shown the critical strange quark mass in a kinematical criterion
		which can be extracted from our result for the unperturbed octet gap as
		$m_s^c(\mu)=2\sqrt{\mu\Delta_8(\mu)|_{m_s=0}}$,.
Suppose that the effective strange quark mass $m_s(\mu)$ is decreasing function of the chemical potential $\mu$.
If $m_s(300~{\rm MeV})>m_s^c(\mu)$ and $m_s(1000~{\rm MeV})<m_s^c(\mu)$, then
		$m_s(\mu)$ intersects the curve of $m_s^c(\mu)$ at some point $\mu_c$ locating 
		between $300~{\rm MeV}<\mu<1000~{\rm MeV}$. 
In the simplified (kinematical) picture, the unlocking transition occurs when the chemical 
	potential $\mu$ gets smaller than this $\mu_c$, where the deviation of 
	the Fermi momentum for strange quark from those for light flavors, 
	$p_F^s-p_F^{u,d}=\sqrt{\mu^2-m_s(\mu)^2}-\mu\sim m_s(\mu)^2/2\mu$,
	exceeds twice of the smallest gap $|\Delta_8|$ in the CFL state.
{Here, we had better mention that if we take into account the dynamical effect of the 
	strange quark mass on the formation of gaps, then the gap of quarks belonging to 
	color-flavor octet splits into 4 different values, and the smallest one $\Delta^8_s$ 
	at the critical point sets the true unlocking critical mass $m_s^{c}({\rm true})=2\sqrt{\mu\Delta^8_s}$.
	In the Ref.~\cite{unlock_AR}, the deviation of the smallest gap from the unperturbed
	octet gap $\Delta_8$ is about $20$\% at $\mu=400$ MeV. 
	Thus using the simple kinematical criterion, we might overestimate the critical mass 
	about $100(1-\sqrt{0.8})/\sqrt{0.8}\sim 10$\% of the real value for $\mu=400$ MeV.
}

In the following, we assume that this criterion is useful even for 
		finite temperature, and we put the further ansatz that the effective mass of 
		the strange quark at chemical potential $\mu$ can be approximated to 
		constant $m_s(\mu)=m_s$ for region $300~{\rm MeV}<\mu<1000~{\rm MeV}$.  
Instead of this simplified ansatz, we treat $m_s$ as a parameter,
		change it by hand, and discuss the qualitative dependence of the phase diagram on $m_s$. 

\subsection{Phase diagram and its $m_s$-dependence}
\begin{figure}[tbp]
  \centerline{
  \epsfsize=0.7\textwidth
  \epsfbox{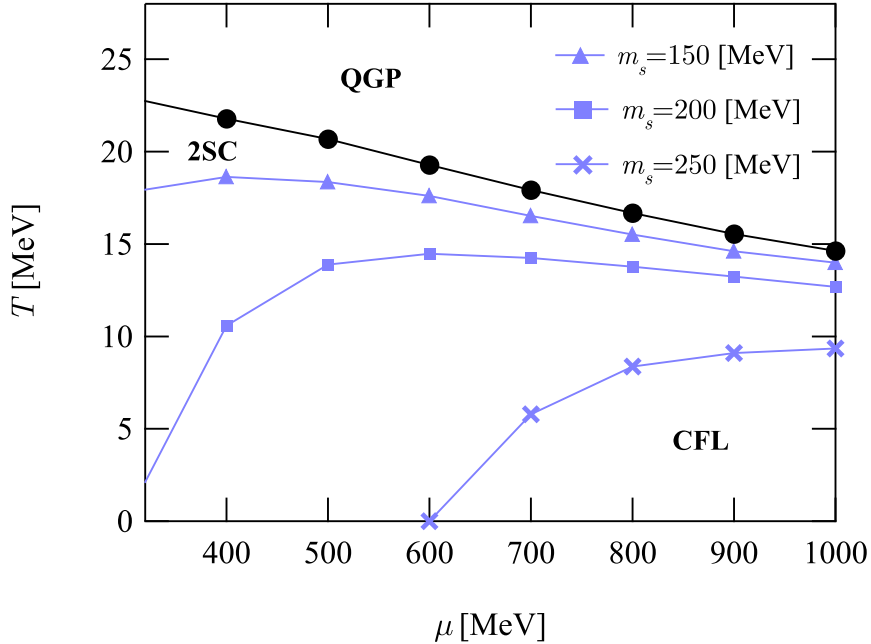}}
 \caption[]%
	{The QCD phase diagram for various values of the strange quark mass. 
						The bold-dot-line represents the phase boundary between 
						2SC and QGP separated by the 2nd order phase transition. 
						Triangle-line, square-line and cross-line are showing
						the gradual change of the unlocking critical line
						when we shift the strange quark mass as $m_s=150,~200$ and 
						$250$~MeV by hand.}
\label{fig:final}
\end{figure}

Here, we draw a phase diagram expected to be realized from $\mu\sim300$~MeV 
		to $\mu=1000$~MeV for various values of the strange quark mass. 
As already noted above, we consider constant effective strange masses without $\mu$-dependence.
Fig.~\ref{fig:final} shows our result of the QCD phase diagrams for four values 
		of the strange quark mass $m_s=0,~150,~200$ and $250$~MeV.

\vspace*{0.2cm}
\noindent
{\it 1.~The chiral limit $m_s=0$~:~~}In 
the chiral limit $m_s=0$, there is only one line (black bold dot) which divides 
	the $(\mu,T)$ plane into the QGP and CFL phases by the 2nd order transition.
This degenerates in double lines, namely, one separating the QGP and the CFL, 
		and the other dividing the QGP and the 2SC.
However as we discussed above, the 2SC phase does not appears for $m_s=0$,
		because the condensation energy of the CFL is about twice larger than
		that of the 2SC for all region $T<T_c(\mu)$.

\vspace*{0.2cm}
\noindent
{\it 2.~$m_s=150$~MeV~:~~}We 
study what happens if we switch on the strange quark mass $m_s=150$~MeV.
We estimate this using the kinematical criterion descried at the end of  the previous section.
New line denoted by triangle-line appears in the phase diagram.
We call this {\it color-flavor unlocking critical line}. 
This unlocking line divides the superconducting region $T<T_c(\mu)$ 
		into the CFL and 2SC phases by 1st order transition.
Above this unlocking critical line, the system is unlocked into the 2SC state, 
		and below it, the system stays in the CFL phase. On the unlocking critical line, 
		the pressure for the CFL coincides with that for the 2SC.
The strength of the 1st order phase transition gets weaker and weaker towards $\mu\to\infty$,
		because the gap between energy density for the 2SC phase and that for the CFL phase 
		gets smaller as one going near the unlocking line.
The critical end point at which the 1st order transition terminates is located at $\mu=\infty$
		as long as we use a kinematical criterion, but the dynamical effect of the strange quark mass 
		on the pairing may bring this point to the finite density point.
{Indeed, in the NJL model analysis \cite{buballa}, the tricritical point appears
		at $\mu\sim 520$ MeV, although the physical reason of the appearance 
		of 2nd order phase transition has not yet been understood. 
	Furthermore, because this tricritical point is quite close to the cut-off 
	$\Lambda\sim 600$ MeV, it might be an artifact of such an effective model 
	with a cut-off scale.
}

\vspace*{0.2cm}
\noindent
{\it 3.~$m_s=200$~MeV~:~~}Now, 
we change the strange quark mass $m_s=150$~MeV to $m_s=200$~MeV.
Then the unlocking critical line shifts from the ``triangle-line''
to the ``square-line''. The room for the 2SC realization is enlarged,
pushing the CFL area to higher density and lower temperature region.
Here we assume the chiral transition occurs near $\mu=400$~MeV,
then at zero temperature, the system is always chirally broken by
the $q\bar{q}$ condensate for $\mu<400$~MeV, and by the CFL type
$qq$ condensate for $\mu>400$~MeV. In this situation $m_s<200$~MeV, 
quark-hadron continuity might possibly realize.

\vspace*{0.2cm}
\noindent
{\it 4.~$m_s=250$~MeV~:~~}Finally, 
the ``cross-line'' represents the unlocking critical line
for $m_s=250$~MeV. In this case, the 2SC might intrude into
the opening region between the chirally broken vacuum phase
and the CFL phase in $T=0$ section in the QCD phase diagram.
Quark-hadron continuity does not occur for $m_s>250$~MeV in our model.
{The critical mass for the quark-hadron continuity lies in the interval
	$200<m_s^{\rm cont}<250$ MeV in our model. The dynamical effect of the strange
	quark mass on the pairing gap would bring this mass to lower densities.
 	The true critical mass $m_s^{\rm cont}$ from our model would be a little bit smaller 
	than $\sim 250$ MeV obtained from the NJL model in Ref.~\cite{unlock_AR}.
}

\section{Summary and Discussion}\label{sec:4}
We have applied the Schwinger-Dyson equation in the improved ladder
		approximation to the color superconducting phenomena in high density quark matter. 
Special attention has been given to the competition between 
		the 2SC phase and the CFL phase at finite temperature,
		and to the effect of the strange quark mass on the QCD phase diagram.

Sec.~\ref{sec:2} was devoted to the technical details to obtain the gap 
		equations and the CJT potentials for CFL and for 2SC from the 
		Nambu-Gor'kov Schwinger-Dyson approach.
In Sec.~\ref{sec:3}, we studied which ordering of (2SC,~CFL) is favored at 
		relatively low density $\mu=300\sim 1000$~MeV at finite temperature. 
		We have drawn the picture of the phase transition by investigating  the $\mu$/$T$ 
		dependence of the gap function, 	the correlation function and the coherent length, 
		as well as bulk quantities such as pressure.
		We summarize our main results in this paper below.

\vspace*{0.2cm}
\noindent
{(1)}~%
The phase transitions to the normal phase both from the 2SC and from the CFL
		have shown almost the same characteristic behaviour as in the BCS weak coupling
		superconductor with contact interaction. 
The critical temperatures for the CFL and the 2SC coincides with each other due to 
		the color anti-triplet dominance, despite the fact that the condensation energy density is 
		almost twice larger in the CFL than in the 2SC.
This facts are consistent with weak coupling estimates, and are indicating that the CFL 
		state is more fragile than the 2SC state against thermal fluctuations, due to the 
		smaller energy gap in the color-flavor octet channel. 
Phase transition occurs as a consequence of the reduction of the coherent Cooper pairs,
		but not due to the dissociation of those.

\vspace*{0.2cm}
\noindent
{(2)}~%
The strong coupling effects get larger towards the lower density side.
Especially, this was observed in the violation of the universal relation between
		the gap at $T=0$ and the critical temperature. 
This value deviates from the weak coupling analytical value $\sim 15$\% at $\mu=300$~MeV.
This deviation comes from strong coupling effects, the large off-Fermi surface
		effect including the pairing in the antiquark channel owing to a larger coupling at lower densities.
We find that physically interesting densities lie far away from the region where this
		ratio seems to converge to the universal value.
Despite these strong coupling effects, the ratio of the physical quantities
		in the CFL state and in the 2SC state, does not so much differ from
		the weak coupling value even in the strong coupling region.

\vspace*{0.2cm}
\noindent
{(3)}~%
The QCD phase diagram and the effect of the strange quark mass $m_s$ on it are studied within 
		a simple unlocking criterion.
In the chiral limit, there is no room for the 2SC realization, but the finite strange quark mass 
		makes a unlocking critical line separating the superconducting $(\mu,T)$ area into the 2SC phase
		and the CFL phase, 
		and this line push the CFL region to higher density and lower temperature region. 
The critical end point on which the 1st order unlocking line terminates is located at $\mu=\infty$
		as long as a simple kinematical criterion is used. 
We find a possibility that for $m_s<200$~MeV, the chiral broken vacuum phase might 
		continuously connected with the CFL phase at high density without any phase transition.

\vspace*{0.2cm}
There are several future problems to be studied.

{\it Dirac mass function and chiral condensate~:~ }In 
order to study the phase boundary region where the chiral 
			restoration occurs, we have to include the diagonal self-energy 
			in our SD equation. Within the 4-Fermi contact interaction model, 
			the competition between the chiral condensate and diquark 
			condensate is investigated by several authors \cite{Berges-Rajagopal,Diakonov}.
			Also, the SD approach is examined in Ref.~\cite{Takagi} for $m_s=\infty$.
{In particular, the $s\bar{s}$ condensate, or equivalently, the dynamical strange 
			quark mass is shown to play an important role
			for the phase boundary between the 2SC and the CFL \cite{unlock_SW},
			which strongly stabilizes the 2SC phase with finite strange quark mass.
			The critical strange quark mass for unlocking transition may be significantly
			lowered by this effect \cite{unlock_SW,buballa}.
}
			
{\it Wavefunction renormalization and chemical potential renormalization~:~~}There 
is an infrared singular factor in the fermion wavefunction
			renormalization : $Z^{-1}(k_0)=1+{\rm const.}g^2\log(\mu/q_0)$ 
			at the one loop level\cite{Holstein}.  Furthermore, the chemical potential
			also suffers from a renormalization because if $\mu\neq 0$, Lorentz symmetry
			is explicitly broken down to $O(3)$. The full self-consistent treatment 
			of both the gap, $Z^{-1}$ and a  chemical potential renormalization $\mu_{\rm r}$ 
			is needed for more quantitative argument.

{\it Exact treatment of the strange quark mass~:~~}In 
this paper, the effect of the strange quark mass was examined within a simple 
			kinematical criterion. 
	It would be interesting to examine the full treatment at finite temperature in our 
			improved SD approach \cite{Inprogress}.
	So far, the unlocking phase transition at finite temperature has not yet studied in 
			a perfect way.
{Including the dynamical effect of the strange quark mass on the gaps, 
			we can make sure whether 
			the critical end point might show up at finite chemical potential or not,
			namely the end-point obtained in Ref.~\cite{buballa} is an artifact
			of an effective model with cut-off or not.
}

{\it Electric charge and color neutrality~:~~}Imposing the color and the
	electric charge neutrality has been shown to strongly disfavor the 2SC 
   by the model independent analysis by Alford and Rajagopal \cite{Alford_CN}.
	Their argument based on the weak coupling expansion of thermodynamic potential 
	in $\Delta\sim m_s^2/\mu$ reveals that the 2SC state hardly realizes in the neutron 
	stars. Several works \cite{CN_NJL,2SCCN1,2SCCN2} also show that the naive 2SC 
	phase costs the large energy for enforcing the neutrality.
	It would be interesting to examine our improved SD approach to the neutral
   quark matter, and to determine the phase diagram under neutral condition.

{\it Another possibility of the ground state~:~~}As 
long as we restrict ourselves to the homogeneous system, 
			the unlocking transition is unique for a given chemical potential $\mu$,
			and thus the $m_s$ dependent unlocking critical line appears in the phase diagram.
			However, if we remove this restriction, then this line might split into two lines
			and another inhomogeneous state like the LOFF state\cite{LO,FF} might 
			intrude between the CFL and the 2SC. For the 2SC phase with the chemical potential 
			difference $\delta\mu=\mu_u-\mu_d$, this phenomena has been investigated 
			in Ref.~\cite{raja_loff}. 
			Recently, it has been suggested that the interior gap state may overcome the LOFF 
			state if the coupling between heavy quark and light quark exceeds some critical 
			coupling\cite{InteriorGap}.
			It would be interesting to study these other possibilities than the pure BCS state
			in our SD approach with the momentum-dependent improved coupling.

\section*{Acknowledgments}
I am grateful to T.~Hatsuda for fruitful discussions and suggestions
		during this work, and for reading of the manuscript.
I wish to express my gratitude to T.~Kunihiro for helpful advices
		on summarizing this paper at the last stage of the work.
I would like to thank A.~Hayashigaki, S.~Sasaki and K.~Itakura 
		for their useful comments and encouragements.

\end{document}